\documentclass{aa}
\usepackage{dblfloatfix}

\usepackage{graphicx}
\usepackage{txfonts}
\usepackage{comment}
\usepackage{makecell}
\usepackage[%
    colorlinks=true,
    pdfborder={0 0 0},
    linkcolor=blue
]{hyperref}
\usepackage{xcolor}
\bibpunct{(}{)}{;}{a}{}{,} 
\definecolor{commentcolorGB}{RGB}{30,144,255}
\def\noteGB#1{{\small \textcolor{commentcolorGB}{$>$GB #1  $<$GB}}}
\definecolor{commentcolorJPB}{RGB}{50,205,50}

\begin{document} 

    \title{FU Orionis disk outburst: evidence for a gravitational instability scenario triggered in a magnetically dead zone\thanks{Based on observations performed with VLTI/PIONIER under program ID 2102.C-5036(A).} \\[1ex] }

   \author{G. Bourdarot
          \inst{1,2,3},
          J-P. Berger
          \inst{2},
          G. Lesur
          \inst{2},
          K.Perraut
          \inst{2},
          F.Malbet
          \inst{2},
          R.Millan-Gabet
          \inst{4},
          J-B. Le Bouquin
          \inst{2},
          R.Garcia-Lopez
          \inst{5,6}
          J.D.Monnier
          \inst{7},
          A.Labdon
          \inst{8,9},
          S.Kraus
          \inst{9},
          L.Labadie
          \inst{10},
          A.Aarnio
          \inst{11}
          }

   \institute{Max Planck Institute for Extraterrestrial Physics, Giessenbachstr.1, 85748 Garching, Germany
         \and
             Univ. Grenoble Alpes, CNRS, IPAG, 38000 Grenoble, France
         \and
             Univ. Grenoble Alpes, LIPHY, 38000 Grenoble, France
        \and
            Giant Magellan Telescope Organization, 465 N. Halstead St., Pasadena, CA 91107, USA
         \and
            School of Physics, University College Dublin, Belfield, Dublin 4, Ireland
        \and
            Max-Planck-Institut für Astronomie, Königstuhl 17, 69117 Heidelberg, Germany
        \and
            Department of Astronomy, University of Michigan, Ann Arbor, MI 48109, USA
        \and
             European Southern Observatory, Casilla 19001, Santiago 19, Chile
        \and
            University of Exeter, School of Physics and Astronomy, Astrophysics Group, Stocker Road, Exeter, EX4 4QL, UK
        \and
            I. Physikalisches Institut, Universität zu Köln, Zülpicher Str. 77, 50937, Köln, Germany
        \and
            The University of North Carolina at Greensboro, USA \\
        \email{bourdarot@mpe.mpg.de}
             }

   \date{Received: 22 December 2022; Accepted 07 April 2023}
   \titlerunning{Constraining the evolution of FU Orionis outburst with near-IR interferometry}
   \authorrunning{Bourdarot et al.}
 
  \abstract
   {FUors outbursts are a crucial stage of accretion in young stars. However a complete mechanism at the origin of the outburst still remains missing.}
   {We aim at constraining the instability mechanism in FU Orionis star itself, by directly probing the size and the evolution in time of the outburst region with near-infrared interferometry, and to confront it to physical models of this region.}
   {As the prototype object of the FUors, FU Orionis has been a regular target of near-infrared interferometry. In this paper, we analyze more than 20 years of near-infrared interferometric observations to perform a temporal monitoring of the region of the outburst, and compare it to the spatial structure deduced from 1D magneto-hydrodynamic (MHD) simulations.}
   {We measure from the interferometric observations that the size variation of the outburst region is compatible with a constant or slightly decreasing size over time $-0.56^{+0.14}_{-0.36}\,\text{AU}/100\,\text{yr}$ and $-0.30^{+0.19}_{-0.19}\,\text{AU}/100\,\text{yr}$ in the H and K band respectively. The temporal variation and the mean size probed by near-infrared interferometry are consistently reproduced by our 1D MHD simulations. We find that the most compatible scenario is a model of an outburst occurring in a magnetically layered disk, where a Magneto-Rotational Instability (MRI) is triggered by a Gravitational Instability (GI) at the outer edge of a dead-zone. The scenario of a pure Thermal Instability (TI) fails to reproduce our interferometric sizes since it can only be sustained in a very compact zone of the disk $<0.1\,\mathrm{AU}$. 
   The comparison between the data and the MRI-GI models favours MHD parameters of  $\alpha_\mathrm{MRI}=10^{-2}$, $\mathrm{T}_{\mathrm{MRI}}=800\,\mathrm{K}$ and $\Sigma_{\mathrm{crit}}=10\,\mathrm{g.cm}^{-2}$, with more work needed both on observations and modeling to precise these values. Locally in the very inner part of the disk, TI can be triggered in addition to MRI-GI, which qualitatively better match our observation but is not strongly constrained by the data available currently.
   The scenario of MRI-GI could be compatible with an external perturbation enhancing the GI, such as tidal interaction with a stellar companion, or a planet at the outer edge of the dead-zone.
   }
   {The layered disk model driven by MRI turbulence is favoured to interpret the spatial structure and temporal evolution of FU Orionis outburst region. Understanding this phase gives a crucial link between the early phase of disk evolution and the process of planet formation in the first inner AUs.
   }


   \maketitle
%

\section{Introduction}

Young stellar objects (YSOs) are surrounded by circumstellar disks, from which they accrete matter in their early phase. In the case of low mass YSOs ($M<3\, \mathrm{M}_{\odot}$, T Tauri stars), it is now clear that accretion should occur through episodic events rather than a steady process \citep{Calvet2000}, as the steady accretion rate deduced from T Tauri is not sufficient to build up a star over timescales of a few million years \citep{Kenyon1990}. Two families of YSOs gather such episodic events : EXor and FUor classes, the archetypical object of the later being the star FU Orionis. The main characteristic of FUors is to display a 4-6mag luminosity outburst at visible wavelengths over short timescales, from few months to few years, followed by a decay in luminosity, from several decades to hundreds of years \citep{Herbig1977}. A large fraction of stars may thus go through FUor/EXor, possibly several times during their early existence. \\
The first comprehensive view of FUors phenomenon has been provided by \cite{Hartmann1996}, and is well explained within the frame of an accretion disk experiencing a massive instability episode, leading to a huge increase of accretion rate, from $10^{-7}\text{M}_\sun\,\text{yr}^{-1}$ to $10^{-4}\text{M}_\sun\,\text{yr}^{-1}$, at the origin of the burst of luminosity. Spectroscopic observations show the presence of an infrared excess in the Spectral Energy Distribution (SED) and of rotationally broadened absorption lines in FUors, giving convincing evidences in favor of a self-luminous accretion disks in rotation around the star. \\ 
Nevertheless, the exact instability mechanism at the origin of the outburst still remains poorly understood, and different classes of solution are generally invoked \citep{Audard2014}. Thermal instability \citep[TI,][]{Bell1994}, in which the outburst is triggered by a thermal runaway of the disk as its temperature reaches ionized hydrogen temperature, was the first mechanism proposed \citep{Bell1994,Hartmann1996}, with discussions on the origin of the trigger of the instability, including external perturbation by an stellar companion \citep{Bonnell1992}, or internal perturbation by the presence of a planet \citep{Lodato2004}. The second possible mechanism is a limit cycle driven by the activation of the Magneto-Rotational Instability (MRI) \citep{Balbus1998} when the ionisation fraction becomes sufficient \citep{Gammie1996}.  The MRI scenario is known affect larger disk regions because it starts at lower temperatures than TIs. It however usually requires an external trigger, which is often assumed to be the Gravitational Instability \citep[GI,][]{Armitage2001} that starts when enough material has accumulated in the outer disc. This scenario was in particular confronted to observational data (SED) in \cite{Zhu2007,Zhu2009}, which favoured this model. 
The last main scenario for the instability is a disk fragmentation \citep{Vorobyov2005,Dong2016}, in which large fragments migrate inwards as the deeply embedded disk becomes gravitationally unstable. \\
FUOrs outbursts are thus intimately related to the process driving accretion in the disk. With that respect, they are particularly important manifestations to understand the mechanisms at the basis of accretion and turbulence in this early phase of a protoplanetary disks, and to characterize the physical properties and structure of the disk. These reasons strongly motivate a comprehensive understanding of the origin of FU Orionis instability mechanism. \\
In this study, we focus on FU Orionis star itself ($d=407\,\mathrm{pc}$ \citep{GaiaEDR3}. The only technique able to spatially resolve the inner astronomical units (AU) of FU Orionis, where the outburst takes place, is near-infrared interferometry. As a relatively bright object in the near-infrared, FU Orionis has been a prime target of interferometry, and was the first YSO to be spatially resolved \citep{Malbet1998}, which gave a direct evidence of the presence of an accretion disk. This initial study was then confirmed and refined in \cite{Malbet2005}, which provided a first estimation of the temperature profile of the disk, and explored the potential detection of an embedded hot spot. This estimation was recently complemented by \cite{Labdon2020}, based in particular on J-band interferometric data on CHARA/MIRC-X, which confirmed the $T\propto r^{-3/4}$ temperature profile of the accretion disk. In addition, the presence of an extended halo as a general feature of FUors in interferometric data, corresponding to the remnant of the infalling envelope, was first identified by \cite{Millan-Gabet2006} on a sample of three FUors. The properties of the dusty envelope of FU Orionis itself was examined in the mid-infrared by \cite{Quanz2006}, and more recently by \cite{Liu2019} through ALMA and GRAVITY data, both of which inferred the presence of a cold envelope.\\
In this paper, we propose to take advantage of the unique temporal coverage accessible with 20 years of archival near-infrared (NIR) interferometry, complemented with dedicated VLTI/PIONIER (ESO run ID 2102.C-5036(A) and VLTI/GRAVITY observations, to perform a temporal monitoring of the outbursting region of FU Orionis. The primary goal of this work is to put constraints on the typical size and variation of the outbursting region in time and to confront this evolution to the spatial structure deduced from 1D magneto-hydrodynamic simulation, in order to identify the mechanism at play in FU Orionis and potentially add preliminary constrains on its physical parameters.
Section \ref{sec:observations} describes the data set used in our study. The global geometrical modeling of FU Orionis, as well as the methodology used to analyze our data sample and give robust estimate of the quality and errorbars is presented in Section \ref{sec:methodology}. The results of this analysis are presented in Section \ref{sec:results}. Section \ref{sec:discussion} provides a comparison of these results with the output of MHD simulation and discusses the implication on the instability mechanism at play in FU Orionis.

\section{Observations}
\label{sec:observations}

\begin{table*}[h!]
\label{tab:log}
\centering                          

\resizebox{\textwidth}{!}{
\begin{tabular}{p{30mm} c c c c c c p{18mm}}        
\hline\hline                 
UT Date & Instrument & Label & Telescopes & Stations & Filter & $R=\lambda/\Delta\lambda$ & Baseline [m] (max) \\    
\hline \\                       
2019/10/03 & MIRC-X & MIRCX19 & 5 & S1-S2-E1-W1-W2 & H & 50 & 313 \\     
2019/02/20 & PIONIER & PION19 & 4 & A0-G1-J2-J3 & H & 30 & 132  \\
2018/11/27 & MIRC-X & MIRCX18 & 6 & S1-S2-E1-E2-W1-W2 & H & 50 & 313  \\
2017/12/25 & PIONIER & PION17 & 4 & D0-G2-J3-K0 & H & 5 & 104\\
2010/12/03 & PIONIER & PION10 & 4 & E0-G0-H0-I1 & H & 5 & 60 \\
1999/11/23 \newline to 1999/12/01 & PTI & IOTA+PTI & 2 & NS & H & 5 & 110\\
1998/12/13 \newline to 1998/12/26  & IOTA & IOTA+PTI & 2 & S15-N15,  S15-N35 & H & 5 & 38 \\
\hline                                   
\hline \\
2021-01-09 & GRAVITY-SCI & GRAVI21 & 3 & U1-U3-U4 & K & 4000 & 130 \\
2017-02-02 & GRAVITY-FT & GRAVI17 & 4 & A0-G1-J2-K0 & K & 20 & 132\\
2016-11-25 & GRAVITY-SCI & GRAVI16 & 4 & K0-G2-D0-J3 & K & 500 & 104 \\
2011-10-27 & CLIMB & CLIMB11 & 3 & S2-E2-W2 & K & 5 & 330 \\
2010-11-29 & CLIMB & CLIMB10 & 3 & S1-E1-W1 & K & 5 & 330 \\
2003/11/19 \newline to 2003/11/27 & PTI & PTI03 & 2 & NS, SW & K & 5 & 110 \\
2002/10/28 & VINCI & VINCI02 & 2 & UT1-UT3 & K & 5 &  \\
2000/11/18 \newline to 2000/11/27 & PTI & PTI00 & 2 & NS, NW & K & 5 & 102\\
1999/11/23 \newline to 1999/12/01 & PTI & PTI99 & 2 & NS & K & 5 & 110 \\
1998/11/14 \newline to 1998/11/27 & PTI & IOTA+PTI & 2 & NS & K & 5 & 110 \\
1998/12/13 \newline to 1998/12/26 & IOTA & IOTA+PTI & 2 & S15-N15, S15-N35 & K & 5 & 38 \\


\hline

\end{tabular}}
\caption{Logs of FU Orionis interferometric observations in H and K Band. The data for which the data-reduction was done in this study are indicated with '\textit{reduced}'. }
\label{table:FUOrionis_log}      
\end{table*}

We based our analysis on dedicated observations performed with PIONIER and GRAVITY instruments at VLTI, obtained in February 2019 (ESO run 2102.C-5036, PI:G.Bourdarot) and January 2021 (ESO GTO run 106.212G.004) respectively, and of the archival data available for these instruments. In addition, we benefited from the broad archive of observations in the H and K band, carried out from 1998 to 2005 and initially published in \cite{Malbet2005}. These data were complemented with CHARA observations published in \cite{Labdon2020}, acquired with CLIMB instruments in K band and MIRC-X data in H band.
Given the heterogeneity of the dataset, a specific emphasis was put on the examination of the data quality, in order to identify potential sources of bias for each observation. The different informations relative to the data as well as the acronym by which they will be designated in the following are provided in Tab.\ref{tab:log} . In the following, we give a description of the different instruments, data sets, as well as reduction tools, we used to process homogeneously our data.

\subsection{Interferometric archival data}
The most distant data in our data set originates from the legacy study of \cite{Malbet2005}. This data set is essentially based on IOTA and PTI observations. These observations consist in two apertures observations, at low spectral resolution in H and K band. IOTA and PTI observations cover complementary spatial scales (respectively small baselines $\sim 30\,\mathrm{m}$ and large baselines $\sim 120\,\mathrm{m}$). These observations were initially gathered as a unique temporal point in the original study of \cite{Malbet2005}, in order to obtain the most complete coverage of the (u,v)-plane. In the present study, we have chosen to separate temporally the PTI observations for each year, and when IOTA data were available, to gather them with PTI data of the closest year in order to benefit simultaneously from constraints on small and high spatial scales. Given the specific format of these observations, the reduced data included in this analysis were provided by Regis Lachaume (private communication). 

In addition to IOTA and PTI observations, we have included CHARA/CLIMB three-telescopes observations (PI : R.Millan-Gabet) which were conducted on 2 nights on UT2010-11-29 and UT2011-10-27, and initially published in \citep{Labdon2020}. Due to the larger number of telescopes and the larger baselines, these observations offer a more complete sampling of the (u,v)-plane in a single observation. In the case of CLIMB10 data, a large dispersion can be seen in the visibility data, as well as a significant discrepancy at high spatial scales (low visibilities) for the same baselines with respect to CLIMB11 data. We will keep CLIMB10 and CLIMB11 in our sample in the following, as they enable to constrain the disk around year 2010 in the K band, however a higher level of confidence will be given on CLIMB11 in the analysis given its lower dispersion in visibility data. These observations were reduced using J.D.Monnier's pipeline at University of Michigan. 

Finally, we also include two CHARA/MIRCX six-telescopes observations, conducted on UT2018-11-27 and UT2019-10-03, and initially published in \cite{Labdon2020}. The data were reduced using the MIRC-X standard reduction pipeline \footnote{https://gitlab.chara.gsu.edu/lebouquj/mircx\_pipeline}. 

\subsection{New data-reduction on archival data}
We also include in our analysis additional archival data, for which it was possible to perform a full new data reduction. 

In the H band, we included two PIONIER observations. PION10 originated from commissioning observations, retrieved from the \texttt{oidb} service of the Jean-Marie Mariotti Center (JMMC) \footnote{http://oidb.jmmc.fr}. PIONIER 2017 data were retrieved from the publicly available ESO archive \footnote{http://archive.eso.org} (ESO run ID 0100.C-0278(J) ). We reduced both datasets using the standard \texttt{pndrs} pipeline \citep{LeBouquin2011}. PION10 shows a relatively large dispersion, but visibility values consistent with other small baselines observations such as IOTA. On the contrary, relatively low data dispersion and small errorbars are seen on PION17 data, but with significantly higher visibility values than similar observations on the ATs with PIONIER. A careful inspection of the calibrators did not show anomalies in the calibration procedure. However, intermediate products of the data reduction highlights strong phase variations during that night, which are likely to bias the visibility of the fringes. These variations could originate from a dome-seeing effect, which would explain such degradation specifically in very good atmospheric conditions. In the following, we included the PION17 data set but we caution that these data may be biased.

In the K band, three additional archival data sets were included. We reduced VINCI observation originally included in \cite{Malbet2005} using the standard \texttt{vndrs} pipeline \citep{Kervella2004}. VINCI data were obtained on the UTs, for which the interferometric field of view is smaller than the ATs given the larger telescope diameter. The extended flux in this data set is thus lower than the other data. We also reduced two GRAVITY datasets on the ATs. For both observations we used the fringe-tracker (FT) channel rather than scientific (SCI) channel of GRAVITY. In the case of GRAVI17, we used the fringe-tracker data only, due to insufficient SNR in the science channel in high-resolution mode ($R\approx4000$). Concerning GRAVI16, a careful inspection of the intermediate product of data reduction highlights discrete phase fluctuations of the fringe tracker away from the zero Optical Path Differrence (OPD) reference. This perturbation potentially impacts the chromatic dependency seen in the visibilities : this point is discussed in more details in Sec.\ref{sec:envelope}.

\subsection{Dedicated interferometric observing programs}
In order to add two contemporary reference observations in H and K band respectively, we performed two dedicated observations of FU Orionis on PIONIER and GRAVITY instruments on VLTI.

The PIONIER observation was obtained through a specific DDT proposal awarded in winter 2019 (PI/Bourdarot), which was conducted on night 2019-02-20 (ESO run 2102.C-5036). PIONIER observation with large configuration on the 1.8m Auxiliary Telescopes (ATs) was specifically chosen in order to reach the highest angular resolution and to obtain the most complete (u,v)-coverage on VLTI. Data were reduced using the standard \texttt{pndrs} pipeline \citep{LeBouquin2011} and exhibit good atmospheric conditions and low statistical dispersion. GRAVITY observations on UTs were obtained on GRAVITY-GTO time (ESO run 106.212G.004) on the 8.2m Unitary Telescopes (UTs) and reduced through the standard ESO pipeline, but this data only includes 3 telescopes due to a technical problem on UT2 during that night.

\subsection{Spectral Energy Distribution}
In order to complement our analysis, we also provide a recent SED of FU Orionis using Gaia DR2, DR3 and TESS photometric data. A particular attention has been drawn to gather data originating from a limited range of time (2015-2019) in order to avoid potential temporal variation. 

\section{Methodology}
\label{sec:methodology}

\begin{table*}[t!]
\label{table:model}      
\centering                          
\begin{tabular}{c c c }        
\hline\hline                 
Component & Visibility Model & Fitted Parameters \\    
\hline                        
Inner-disk & $\exp\left(-\frac{\pi a \sqrt{\alpha^2+\beta^2}}{4\log{2}}\right)$, \quad 
\small $\begin{cases} 
    \alpha =& u\cos\theta-v\sin\theta \\
    \beta =& \Big(u\cos\theta+v\sin\theta\Big)\cos i
    \end{cases}$
& $a, \theta, i$\\ \\
Unresolved component & $f_c$ & $f_c$ \\ \\
Extended disk & $f_e(\lambda)= \left(f_0+f_1(\lambda-\lambda_0)+f_2(\lambda-\lambda_0)^2\right)\cdot \delta(0,0)$ & $(f_0,f_1,f_2)$ \\
\hline 
\end{tabular}
\caption{Geometrical model used to fit FU Orionis visibility.}             

\end{table*}

In this section, we describe the geometrical modeling of FU Orionis and the fitting strategy used for our temporal analysis. The final goal of this analysis is an estimation of the full-width-at-half-maximum (fwhm) size of the outburst in time, together with a reliable estimation of the statistical and systematics errors in the data through bootstrap analysis. 

\subsection{Geometrical modeling of FU Orionis}

In order to introduce the geometrical modeling of FU Orionis, we present one observation which exhibits the archetypal features of our sample. Three main components can be identified in the visibility, that will be designated as the inner disk (partially resolved), the central unresolved component, and the over-resolved structure. This component describes the general structure of FUors object, as shown in Fig \ref{fig:mircx18}. This model is also consistent with the common picture from NIR interferometric observations in Herbig Ae/Be stars \citep{Lazareff2017,Perraut2019}.

\subsubsection{Inner-disk}
The central component of the visibility component visible on the interferometric observations is a regularly decreasing visibility profile, whose extent is inversely proportional to the typical size of the emission zone. This component is marginally resolved on spatial frequencies $45 \mathrm{M}\lambda \equiv 100 \, \mathrm{m}$ in K band, and therefore will be modeled as a gaussian elongated disk $V_{d}(\alpha,\beta)$, with $a$ the full-width-at-half-maximum (fwhm) of the gaussian, $\theta$ the position angle of the disk, and $i$ the inclination angle of the disk (cf Tab \ref{table:model}). 
In the geometrical model, this component will be designated as the resolved emission. In FU Orionis, this resolved component is associated to the inner disk surrounding FU Orionis that constitutes the brightest component of the object, where the instability at the origin of the outburst takes place. In this respect, the primary goal of the present study is to constrain the size of this emitting region as a function of time.

\subsubsection{Extended structure}
\begin{figure}[b]
   \centering
   \includegraphics[width=8.5cm]{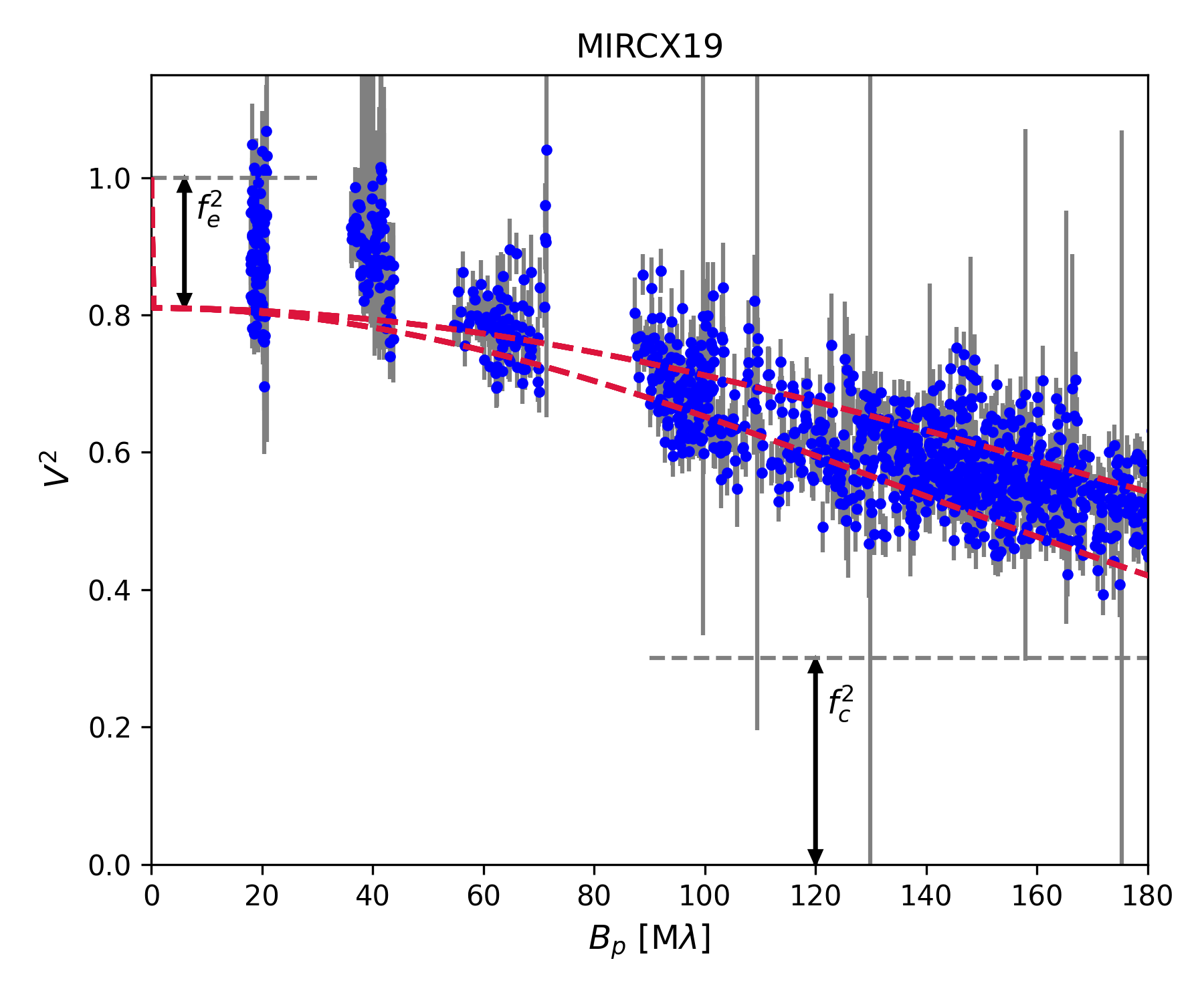}
      \caption{Visibilities squared versus projected baseline, observed with MIRCX instrument in H-band in year 2019 (see logs Tab \ref{table:FUOrionis_log}). Red dashed line shows the minor and major axis of the gaussian inner disk ; black solid lines and black arrows show the contribution of the compact flux $f_c$  and of the over-resolved flux $f_e$.}
         \label{fig:mircx18}
\end{figure}

FU Orionis is embedded in a large extended emission \citep{Millan-Gabet2006}, also visible on high-contrast images at very large spatial scales \citep{Takami2018}. This emission corresponds to the large infalling envelope which feeds the disk in the classical picture of FUors \citep{Hartmann1996}. In the case of interferometric observations, this extended flux fills the whole interferometric field of view, whose extension corresponds typically to the diffraction limit of one telescope, of the order of $250\,\mathrm{mas} \equiv 100\, \mathrm{AU}$ for an ATs in H band. Finally, in the visibility space, the extended component corresponds to a low spatial frequency object, fully resolved in our observations, which will be modeled as a dirac function with a certain amplitude $f_e$, centered on zero spatial frequency. In Fig \ref{fig:mircx18}, this component appears indeed as an apparent offset between the maximum of the resolved visibility and the $V^2=1$ visibility at null spatial frequency. Given the apparent spectral dependency visible in GRAVITY data, first reported in \cite{Liu2019}, a polynomial dependency in wavelength is allowed for this over-resolved component (second order polynomial), with coefficients $(f_0$,$f_1$,$f_2)$.

\subsubsection{Central unresolved component}
The inner disk is marginally resolved on baselines smaller than 100m (spatial frequencies of $40\,\text{M}\lambda$), and is resolved in H band with 300m baselines (spatial frequencies of $180\,\text{M}\lambda$), as shown on MIRC-X observations (see Fig.\ref{fig:mircx18}) in the H band, and CLIMB observations in the K-band. For the largest spatial frequencies, this resolved component left an unresolved component, which appears as a "floor" of visibility at high angular frequencies. This unresolved component corresponds to a compact emission in the inner part of the disk which is not resolved at the resolution of our near-infrared interferometric data, thus encircled in $<0.1\,\mathrm{AU}$ of the disk. In the following, we model this floor as a pure unresolved component in our geometrical model i.e. a constant in visibility $f_{c}$ (see Tab \ref{table:model}).\\

Finally, the total visibility is written as the sum of the three individual components, normalized to 1:

\begin{equation}
    \label{eq:geom_model}
    V(u,v,\lambda)=
    \Big(1-f_c-f_e(\lambda)\Big)\cdot V_{d}(u,v)+
    f_e(\lambda)\cdot\delta(0,0)+
    f_c
\end{equation}

The interferometric data considered in the following will consist in the modulus squared of Eq (\ref{eq:geom_model}).

\subsection{Fitting strategy}

\begin{figure*}[t!]
   \centering   
   \includegraphics[width=19cm]{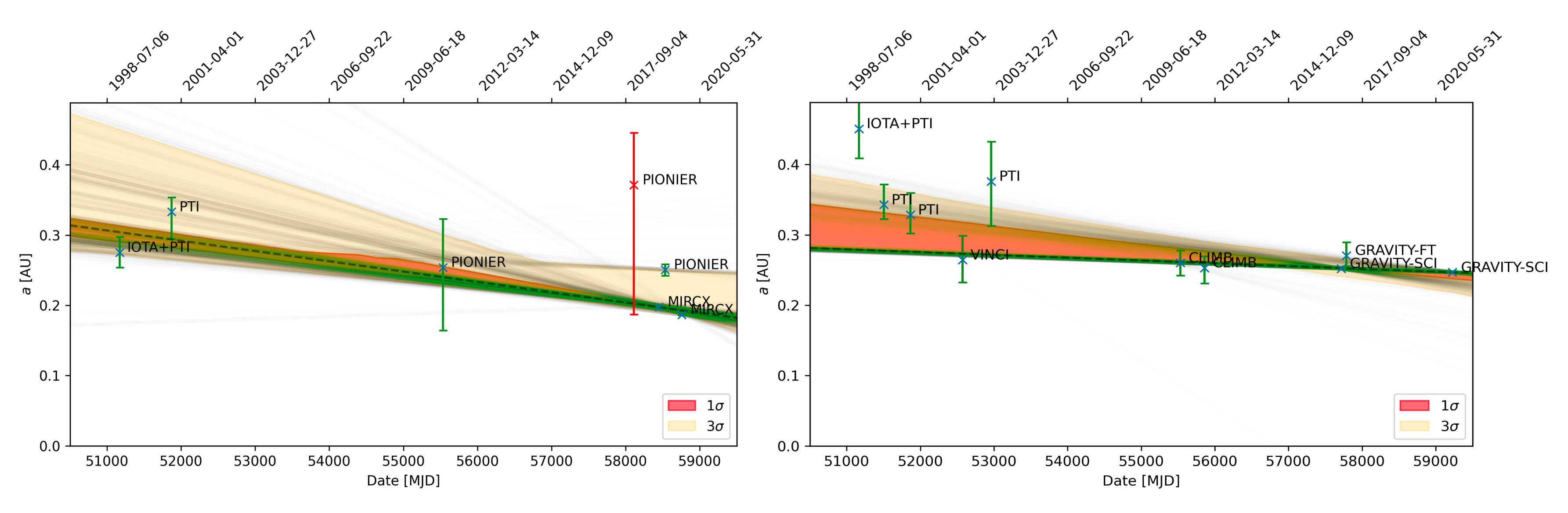}
      \caption{Variation in time of the estimated fwhm size of the outburst in FU Orionis, in H band \textit{(left)} and K band\textit{(right)}. The temporal slope is evaluated from a bootstrap analysis: each trial of the bootstrap is shown in transparent solid line, the final result is the median of all trials, shown in dashed line.}
         \label{fig:temp_slope}
\end{figure*}

The major constraint of this study arises from the large heterogeneity of the dataset, and in some cases from the sparsity of the uv-coverage of the oldest data (IOTA, PTI). These constraints call for a specific fitting strategy. We first present the global methodology used in this study to fit the interferometric data. We then detail this procedure for each particular data set in H and K band. We used a Levenberg-Marquardt minimization over all steps to fit the model to the interferometric data.

Our fitting strategy is decomposed in three steps :\\

\textit{1. Initialization} :
    We start the procedure by fitting independently the whole set of parameters to each observation that present the most complete (u,v)-coverage and the best data uncertainty at medium spectral resolution ($R>50$). These observations are chosen among the second generation instruments (PIONIER and MIRCX in H band, GRAVITY in K band). Then we choose the observation that has the most robust estimate to statistical errors (dispersion) and systematic errors (e.g bi-modal distribution) in the H and K bands. The parameters of this observation will be used as a reference set of parameters in Step 2. for the other, less complete observations. \\
    
\textit{2. Fit of the fixed observations} :
    In the case of the oldest instruments, the complete set of parameters of the visibility model will be poorly constrained due to a limited number of visibility points in the (u,v)-plane and the parameters degeneracy. On the other hand, several parameters of the model are considered constant over time : the position and inclination angles ($\theta$ and $i$), the spectral variation of the envelope $(f_0,f_1,f_2)$, and the flux of the central unresolved component ($f_c$). These parameters are fixed on the reference observations during the fit, and only $a$ is let as a free parameter. \\

\textit{3. Error estimation} :
    The error on the estimation of each parameter is then estimated through the bootstrap method \citep{Efron1982,Kervella2004,Lachaume2019}. For each observation, we draw with repetition a number of visibility points that equals the number of visibility points in the initial observation. We then apply the fit to these points, in the same order than described in Step 1. and Step 2. The fitted parameters are then registered, and the whole procedure (Steps 1 and 2) is repeated from the start over a large number of trials (typically 1000). The final parameters are estimated with the median value of the samples, and the $1\sigma$ upper and lower bar with the $84\%$ and $16\%$ percentile of the distribution. This method also enables us to reconstruct the probability distribution of our fitted parameters and to evaluate their errorbars without any assumptions on their properties. In addition, it is possible to identify from the output probability distribution potential bi-modal or multi-modal results, which could trace potential biases or local minima in our parameter estimation.

The list of the reference and fixed observations, as well as the different parameters fitted for each observations, are given in Table \ref{table:results}. The peculiarities of the fit for each spectral band is detailed in the next subsection.

\paragraph{Fit of the H band data set}~\\
The whole set of parameters of the model are fitted independently in MIRCX19, PION19, MIRCX18 and PION17 (Step 1), considering the (u,v)-coverage and the spectral resolution available with this dataset. A bi-modal distribution can be seen at the output of MIRCX19 bootstrap Fig. \label{fig:outH}, which may originate from an issue with the calibrator, so that we prefered to select MIRCX18 as a reference for the value of $f_c$ ; nevertheless, the values of $f_c$ obtained with MIRCX19 and MIRCX18 are consistent between each other. Given their limited baselines, PIONIER observations barely estimate the contribution of the unresolved component shown in Fig. \ref{fig:mircx18}, so that $f_c$ parameter in PION19 and PION17 are fixed by MIRCX18. PION17 was both tested with independent and fixed parameters, but in both case this fit does not enable to obtain a reliable size estimation, for reasons explained in Sec \ref{sec:observations}. PION10, PTI00 and IOTA+PTI exhibit a sparse (u,v)-coverage, so that their parameters are fixed on a reference observation (Step 2). 
Due to the lack of small baselines, the contribution of the extended envelope in PION10 and PTI00 was fixed on PION19, and only the size $a$ was fitted for these two observations. IOTA observations conducted in 1998 and PTI observations conducted in 1999 (Tab \ref{table:FUOrionis_log}) were combined in the same point IOTA+PTI, as they enable to constrain both small and intermediate baselines respectively, at the price of a relatively low decrease in time resolution, which enables to provide an independent estimation of $a$ and $f_c$. The results of the bootstrap (Step 3) are shown in Tab \ref{table:results}. 

\paragraph{Fit of the K band data set}~\\
In the K band, GRAVITY and CLIMB data provide independent fits of the whole set of parameters. 
The compact component is probed by CLIMB data but is less constrained compared to H band MIRC-X data, due to higher errorbars and a smaller number of visibility points (smaller number of spectral channels and smaller number of baselines). 
For GRAVI21, GRAVI17 and GRAVI16, $f_c$ is not well constrained due to smaller baseline compared to CHARA, so that this value is fixed to the value of fitted with CLIMB ($f_c=0.30$). 
The parameters of PTI observations were fixed on GRAVI17, given the sparsity of these datasets, except $a$ let as a free parameter. GRAVI17 was preferred over GRAVI16 given the fringe-tracking issue in this observation (Sec \ref{sec:observations}), and over GRAVI21 given that the field of view of the ATs are closer to the field of view of PTI and IOTA observations. 
For the same reason, GRAVI21 was chosen as a reference for VINCI data, which were obtained on the UTs. In a similar fashion than data in H band, the IOTA and PTI observations conducted in 1998 were combined in a single point, which allows to provide an estimation of both $a$ and $f_e$. The results of the bootstrap analysis are also shown in Tab \ref{table:results}. 

\section{Results}
\label{sec:results}

In this section, we present the estimation of the parameters through the temporal fitting, and evaluate the range of validity of these results. The complete output of the fit is shown in Appendix, on Fig.\ref{fig:outH}, Fig.\ref{fig:outK_pI} and Fig.\ref{fig:outK_pII} .

\begin{table}[b!]
\centering                          
\begin{tabular}{c c c}        
& H band & K band\\    
\hline                        
\makecell{Mean fwhm\\ $[\mathrm{AU}]$} & $0.23^{+0.04}_{-0.08}$ & $0.28^{+0.03}_{-0.03}$ \\
\hline
\makecell{Slope \\ $[\mathrm{AU}/100\mathrm{yr}]$ } & $-0.56^{+0.14}_{-0.36}$ & $-0.30^{+0.19}_{-0.19}$ \\
\hline 
\end{tabular}
\caption{Estimation of the size and the temporal slope of the resolved emission}             
\label{table:slope}      

\end{table}

\subsection{Size estimate of the inner zone}
The size $a$ of the resolved emission as a function of time is shown in Fig \ref{fig:temp_slope}. On average, the mean size $a$ over all the observations is $0.56^{+0.10}_{-0.20} \,\text{mas}$ in the H band and $0.68^{+0.07}_{-0.07}\,\text{mas}$ in the K band, which are consistent with previous estimates based on comparable models \citep{Labdon2020}. These measurements translate in $0.23^{+0.04}_{-0.08} \,\text{AU}$ in the H band and $0.28^{+0.03}_{-0.03}\,\text{AU}$ in the K band, assuming a distance $d=407\,\mathrm{pc}$ \citep{GaiaEDR3}. 

The error bars and the validity of each individual data point can be inspected through the output of the bootstrap Fig. \ref{fig:outH} and \ref{fig:outK_pI}. The output of the bootstrap follow overall a gaussian distribution in both H and K band, MIRCX19 and PION17 excepted, due to the above-mentioned observational biases. In the following, we will discard PION17, given that this size estimate is clearly inconsistent with the rest of the sample and that an observational bias is identified for this point. From Fig \ref{fig:temp_slope}, the size estimate of the oldest data points in the the early 2000s differ significantly from the most recent points, in particular in K band. On the other hand, the error bars of IOTA and PTI points tends to be underestimated by the bootstrap analysis, given that only two parameters are fitted. 

\begin{figure*}[t!]
\label{fig:model-comparison}
   \centering
   \includegraphics[width=18cm]{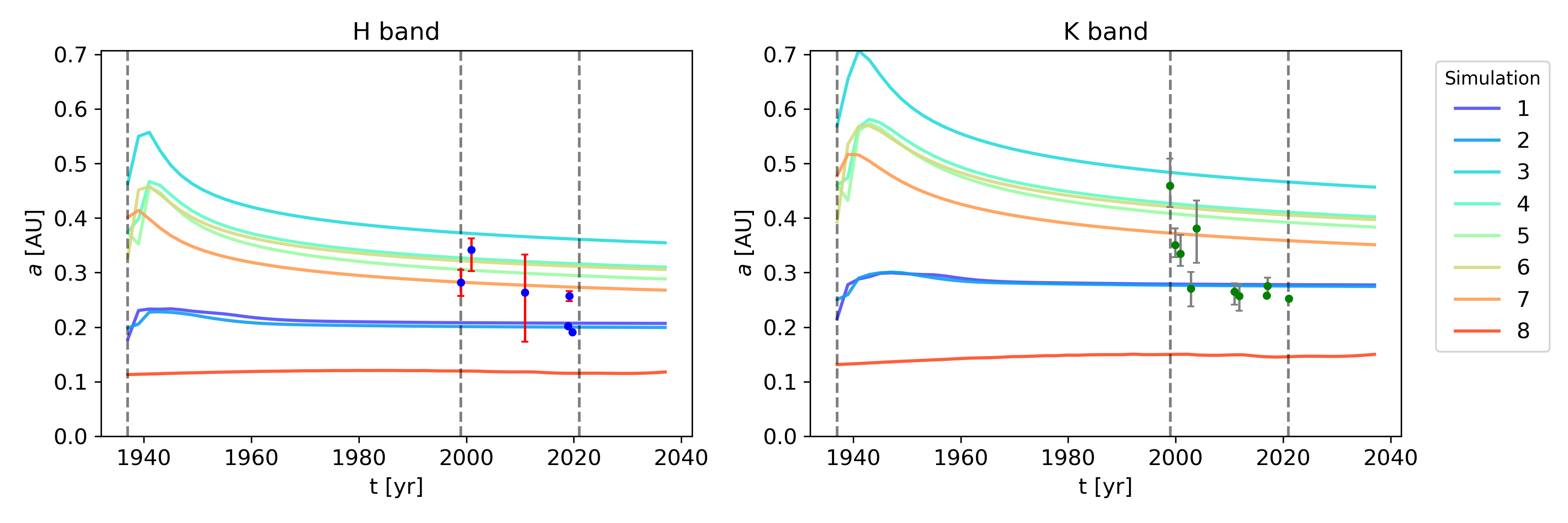}
      \caption{Temporal evolution of the fwhm size of the outbursting region deduced from the model (solid line) compared to the fit of the observations (points). H band observations agree correctly with the model; K band exhibit a good mean estimate but a strong temporal variation for the oldest data points, likely originating in instrumental causes.
      }
\end{figure*}

From this set of fwhm sizes, we estimate the temporal slope in H and K band. 
The temporal slopes resulting from the bootstrap are respectively $-0.56^{+0.14}_{-0.36} \,\text{AU}/100\text{yr}$ and $-0.30^{+0.19}_{-0.19} \,\text{AU}/100\text{yr}$ in H and K band. These two values are consistent with a slowly decreasing or a constant size over time. In addition, the temporal variations are consistent within $1\sigma$ to each other. 
In H and K band respectively, the mean temporal slope is mostly constrained by the points with the smaller uncertainty, i.e. MIRC-X and GRAVITY respectively. Although the uncertainties of the first points in the 2000s are large, they contribute to draw the uncertainty contours of the temporal estimate, and are important to cover a large time span and to set upper and lower bounds on the size variation of the resolved emission. Finally, the overall consistency of this temporal slope with the mean estimate of $a$ can be evaluated by comparing the points lying in the 1$\sigma$ and 3$\sigma$ estimate of the temporal slope in Fig.\ref{fig:temp_slope}. In K band, IOTA+PTI point lies outside the $3\sigma$ region, but might affected by several causes such as the estimation of the extended flux by combining IOTA and PTI data and large error bars in the data. 

\subsection{Spectral variation and temperature of the extended envelope}
\label{sec:envelope}
The over-resolved component was first systematically revealed by \cite{Millan-Gabet2006} as a general feature of FUors observed with single-mode near-IR interferometry. This feature is the signature of a strong extended flux covering the interferometric field of view, in which FUor are still deeply embedded, and which is also clearly visible in AO images. The fit in the K band highlights a spectral dependency of the extended envelope in the GRAVI16 high-resolution data, as reported by \cite{Liu2019}, who analyzed the same Gravity dataset. An uncertainty remained in this dataset since strong OPD shifts could also been seen in the data, which could be confused with the data. We thus also analyzed the GRAVI21 and also observe this spectral dependency in the data, with a total extended flux contribution smaller than GRAVI16. This smaller number is expected since the field of view of the UTs is smaller than the ATS by a factor 20 in surface, which matches the value fitted. The spectral dependency is shown in Fig \ref{fig:envelope}. In addition, in the H band, we do not detect any significant chromatic effect of the envelope in MIRCX data (R=50), which thus tends to indicate a significantly red component.

\begin{figure*}[t!]
\label{fig:global}   \centering
   \includegraphics[width=19cm]{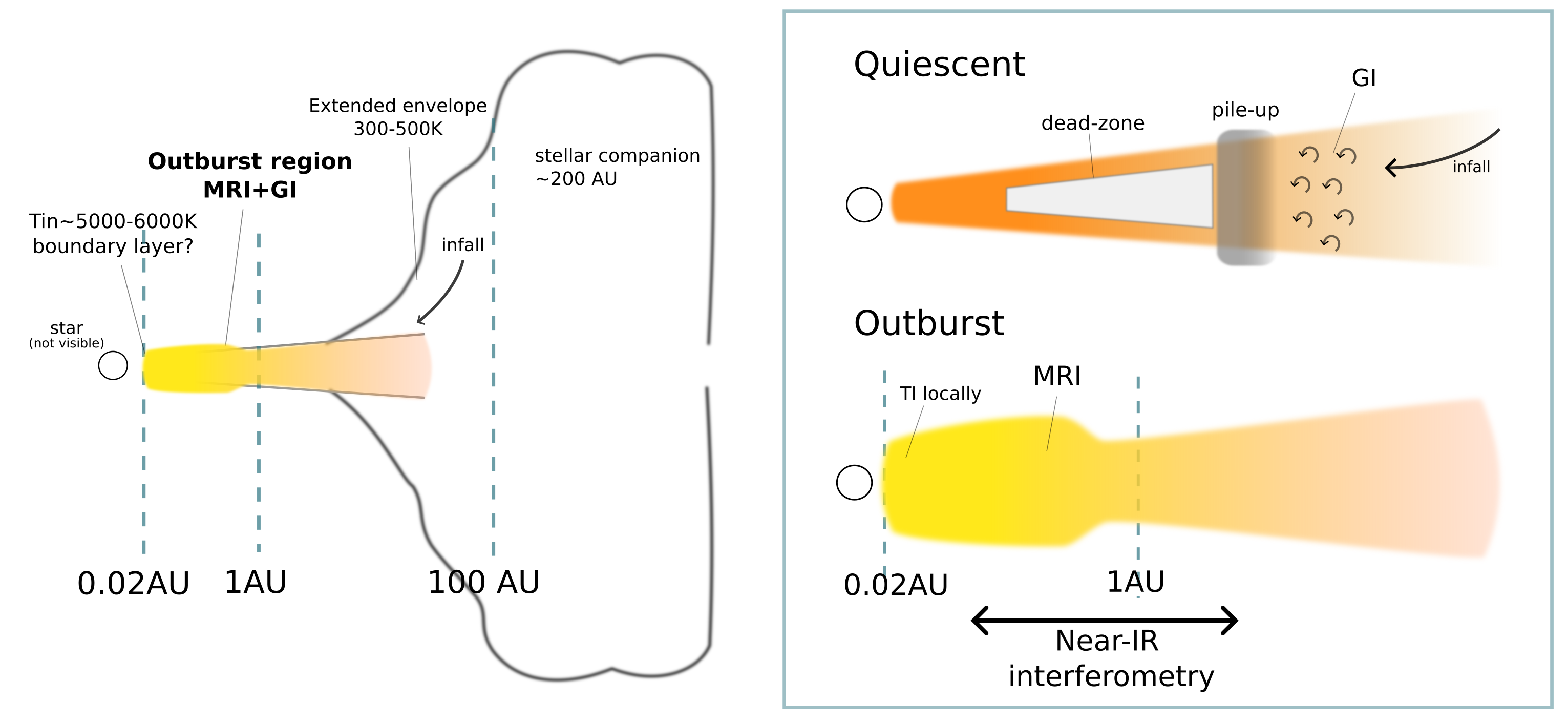}
      \caption{\textbf{Left:} Structure of FU Orionis as seen with the components of our near-infrared data-set. The outbursting region correspond to the marginally resolved disk in the interferometric data. \textbf{Right} (adapted from \citep{Armitage2011}): Scenario of a FU Orionis outburst in the MRI-GI model. In the quiescent phase, the disk has a magnetically layered structure: material piles up at the outer edge of the dead-zone due to Gravitational Instability. As it reaches a sufficient density and temperature to activate MRI, the full disk goes in outburst, which corresponds to the region seen in the near-interferometric data. Locally, Thermal Instability can be triggered very close to the star ($<0.1\,\mathrm{AU}$).}
\end{figure*}

Using both the non-detection in the H band and the spectral slope in the K band, we have fitted a black-body to the spectrum of the envelope, which by definition is the total spectrum multiplied by the proportion of the extended flux $f_e(\lambda)$. We have fitted independently the spectrum of the envelop in GRAVI16 and GRAVI21 in order to estimate the error of this estimate. We obtain respectively a temperature of $T_{\mathrm{GRAVI}16}=580\,\mathrm{K}$ and $T_{\mathrm{GRAVI}21}=280\,\mathrm{K}$. This discrepancy can be attributed to the fact that these two observations probe significantly different field of view and to the accuracy with which we estimate the slope of this component in the fit process. We will keep as indicative value an approximate value of 300K-500K for the temperature of the envelope (Fig \ref{fig:global}). This observation could be compare with interferometric observations in the mid-infrared in order to refine this estimation of the temperature of  halo \citep{MATISSE2022}.
 \\

\subsection{Unresolved flux}
\label{sec:uflux}
The unresolved flux of the disk is better estimated in the CHARA/MIRC-X H-band observations. The mean unresolved flux for these observations are $f_{c}=27.8^{+0.3}_{-0.3}\,\%$ and $f_{c}=29.1^{+0.7}_{-0.8}\,\%$ in the H band. This compact unresolved component thus represents a significant proportion of the total flux $\sim 30\,\%$. All this flux should be encircled within a radius smaller than $\sim 0.1\,\mathrm{AU}$ (i.e within the inner disk), according to the maximum spatial frequency measured in MIRC-X observation. This component can naturally arise from a power-law temperature profile, which we discuss in Section \ref{sec:MRI+GImodel}.\\


\section{Discussion}
\label{sec:discussion}

\begin{figure*}[t!]
\label{fig:1DMHD_MRI}   \centering
   \includegraphics[width=19cm]{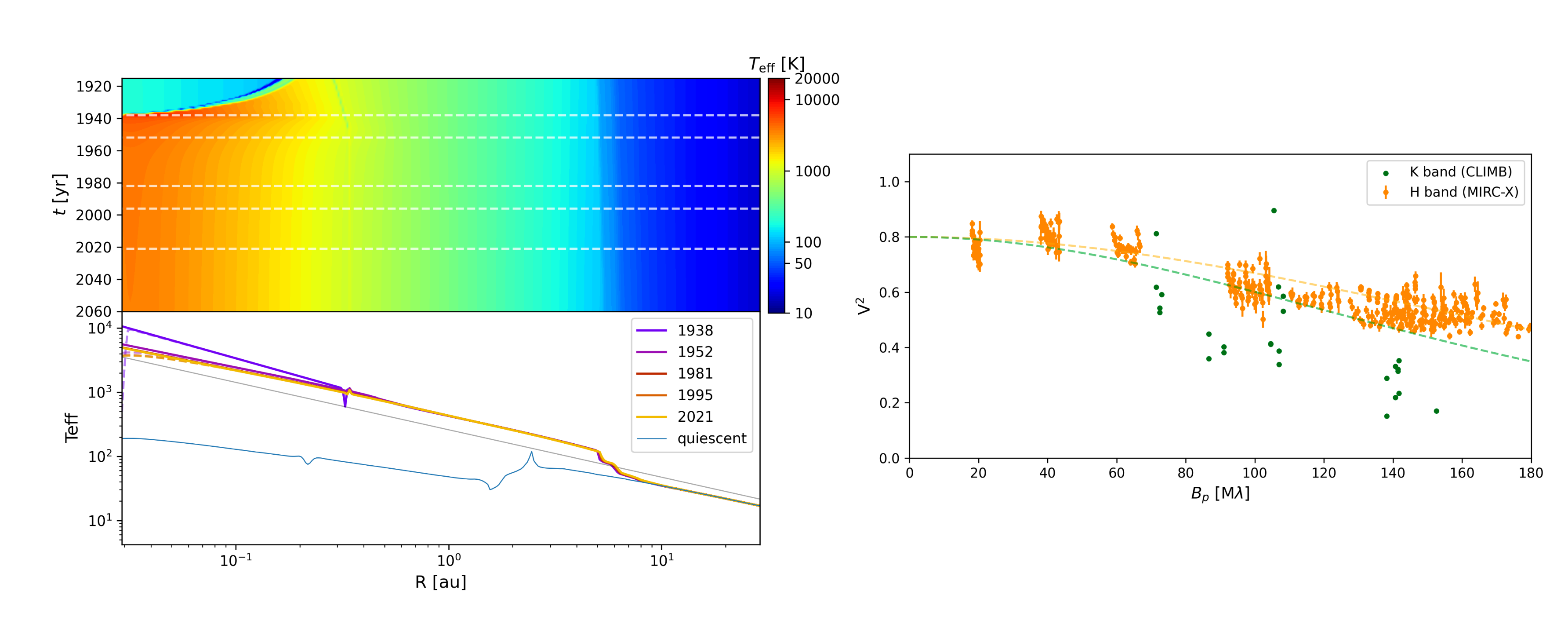}
      \caption{\textbf{Left (upper)}: Simulation of an FU Orionis eruption with a 1D MHD layered disk model (Simu 1) \textbf{Left (lower)}: Temperature radial profile produced by the simulation for different epochs (solid colored lines) ; a $T\propto r^{-3/4} $ slope is shown for comparison (solid gray line) ; the uncorrected profile in the innermost region ($\sim 0.01\,\mathrm{AU}$, see text) is shown in dashed line. \textbf{Right :} The spatial morphology of the disk is then deduced from simulated temperature profile and compared to interferometric observations in H band ($1.65\,\mu\mathrm{m}$) and K band. ($2.2\,\mu\mathrm{m}$).}
\end{figure*}

\begin{table}[b]
\centering                          
\begin{tabular}{c c c c c c c c c}        
 & $T_{\mathrm{MRI}}$ & $\Sigma_{\mathrm{crit}}$ & $R_{\ast}$ & MRI & TI & $r_{\mathrm{in}}$ & $T(r_{\mathrm{in}})$ & $\alpha_\mathrm{MRI}$ \\    
 & [K] & [g.cm$^{-2}$]&  [$R_{\odot}$]& +GI & & [$R_{\odot}$] & [K] &  \\    
\hline                        
1 & 800 & 10 & 5.17 & yes & yes & 5.3 & 5480 & $1.10^{-2}$\\
2 & 800 & 10 & 2.24 & yes & yes & 6.6 & 5250 & $1.10^{-2}$\\
3 & 800 & 10 & 2.24 & yes & no & 12.7 & 5430 & $2.10^{-2}$\\
4 & 800 & 10 & 2.24 & yes & no & 10.9 & 4680 & $1.10^{-2}$\\
5 & 800 & 10 & 5.17 & yes & no & 8.8 & 4900 & $1.10^{-2}$\\
6 & 800 & 40 & 5.17 & yes & no & 10.6 & 4746 & $1.10^{-2}$\\
7 & 600 & 10 & 2.24 & yes & no & 9.0 & 4916 & $1.10^{-2}$\\
8 & - & - & 2.24 & no & yes & 5.0 & 3411 & $1.10^{-2}$\\
\hline 
\end{tabular}
\caption{Parameters of the 1D numerical simulations for the outburst.}             
\label{tab:simulations}      
\end{table}

In this section, we compare our interferometric observations to the outburst models of FUors. Two main models were put forward in the case of FU Orionis \citep{Hartmann1996,Hirose2015} : an outburst driven by Thermal Instability (TI) \citep{Bell1994}, an outburst driven by Magneto-Rotational Instability triggered by Gravitational Instability (MRI+GI) in a magnetically layered disk model \citep{Armitage2001}.

\subsection{Thermal instability model}
The first instability model proposed for FUors is an outburst driven by TI \citep{Bell1994}: in this model, the outburst is triggered as the disk temperature increases up to the point where hydrogen is ionized, creating an abrupt change of opacity, which traps the thermal energy generated by the viscous disk and results in a thermal run-away of the disk. Once the matter has been accreted onto the star, the disk returns to its lower state, until a new cycle begins. We implemented the TI in a 1D numerical model, assuming an axisymmetric disk. The numerical model is based on the 1D disk instability model used by \cite{Scepi19,Scepi20} in the context of cataclysmic variables, omitting magnetised winds. To sample both inner and outer regions, a logarithmic radial grid is used, from $0.1\,\mathrm{AU}$ to 30 AU. We assume a star with a mass $\mathrm{M}=0.5\,\mathrm{M}_{\odot}$ \citep{Hartmann1996,Perez2020}. The opacities used in our model are based on the prescription described in \cite{Bell1994}. Angular momentum transport and accretion are solved in the $\alpha$ disk paradigm \citep{SS73} using a time-dependent continuity equation. In the TI models, $\alpha$ is assumed to have two values: one in the hot state and one in the cold state. The $\alpha$ values have to be as low as $10^{-3}$ in the hot state and $10^{-4}$ in the cold state, in order to be compatible with the typical timescale of the eruption seen in FUors \citep{Bell1994}. However, the exact physical justification of these particularly low values of $\alpha$ remains problematic in general \citep{Hirose2015}. \\

The outburst predicted by TI models alone brings two main difficulties. First, the estimated fhwm of the disk emission from this model is $<0.1\,\mathrm{AU}$, three times smaller than $\sim 0.3\,\mathrm{AU}$ measured in our interferometric observations. This smaller predicted emission arises from the high temperature needed to ionize hydrogen $\sim 5000\,\mathrm{K}$, which can only be sustained in the innermost region of the disk. TI cannot sustain effective temperatures much higher than 5000K up to a maximum extent of $\sim 0.01\,\mathrm{AU}$. Second, without external triggering event, the instability fronts propagates inside-out, starting at few stellar radii where the temperature is high enough and increasing its size as the whole disk becomes unstable, up until the instability front stops its propagation and moves backwards onto the star. This should translate in a increase of size and of the visible magnitude over few tens to one hundred years, followed by a decay over a similar timescale. This contradicts the pure decay in time observed in our interferometric sizes and in the lightcurves. In Sec \ref{sec:simulations}, we present in more details the method to compare these simulations with the observations. \\


\subsection{MRI+GI in a layered disk model}
\label{sec:MRI+GImodel}
The second model considered in this work is a magnetically layered disk driven by MRI-GI \citep{Armitage2001,Armitage2011}. In the quiescent state, the disk is structured with a thin surface layer, in which accretion by MRI can be sustained with the ionization from high-energy particles, and a dead zone from 0.1AU to 2AU \citep{Gammie1996}, as shown in Fig \ref{fig:1DMHD_MRI}. As a magnetically inactive region, the dead zone cannot sustain MRI and is quiescent. On the opposite, the outer region of the disk is considered as gravitationally unstable for radii larger than $\sim 3\,\mathrm{AU}$. Material from the outer of the disk thus flows inwards and piles up on the outer edge of the dead zone. As material accumulates, the temperature increases, up to a point where it reached the activation temperature of MRI ($T_{\text{MRI}}$), thus triggering the instability. This instability front propagates inwards and place the entire disk in a hot turbulent state. After the outburst, as the matter has been drained from the disk, the inner disk returns to its initial quiescent state, until a new cycle is triggered. 

The main difference between the TI and the MRI+GI scenarios  is the threshold temperature for the outburst. In the TI case, the outburst is due to the sudden rise of the opacity due to the ionisation of hydrogen atoms, which starts at about 3000 K. In the MRI case, it is the ionisation of alkali metals (above 800K) that is sufficient to increase the ionisation fraction above that required for MRI activity \citep{Gammie1996}. The net result is that MRI outbursts start at larger radii (lower temperature), and therefore concern a wider fraction of the disk than TI outbursts. Moreover, since the typical accretion rates are comparable in both kinds of outbursts, MRI outbursts last longer than TI outbursts since it takes more time to empty the larger MRI-unstable disk.

Note however that both outbursts can exist simultaneously: TI could locally be activated in the hottest and innermost region of the disk $<0.1\,\text{AU}$, although it is not essential in triggering and sustaining the outburst \citep{Armitage2001}. We implemented the MRI+GI model in a 1D numerical simulation similar to the TI model (axisymmetric, logarithmic scale), based on the model given in \cite{Martin2011}. 

In contrast to the TI model, $\alpha$ is in this case computed as a sum of several contributions:
\begin{align}
\alpha=&\alpha_\mathrm{DZ}+\alpha_\mathrm{AZ}+\alpha_\mathrm{SG}
\end{align}
where we have identified three regimes of angular momentum transport: the dead zone, where the MRI is active only in the surface layer down to a column density $\Sigma_\mathrm{crit}$ \citep{Lesur22}:
\begin{align}
\alpha_\mathrm{DZ}=\alpha_\mathrm{MRI}\frac{\Sigma_\mathrm{crit}}{\Sigma},
\end{align}
the active zone where the MRI is activated in the entire disk column when $T\gtrsim T_\mathrm{MRI}$
\begin{align}
\alpha_\mathrm{AZ}&=\alpha_\mathrm{MRI}\frac{\Sigma-\Sigma_\mathrm{crit}}{\Sigma}\tanh\big[(T-T_\mathrm{MRI})/150\big]\quad\textrm{if $T>T_\mathrm{MRI}$,}\\ &=0\quad \mathrm{otherwise},
\end{align}
and finally the self-gravitating regime, where angular momentum is driven by GI when the Toomre $Q$ parameter \citep{Toomre64} gets below a threshold $Q_\mathrm{crit}$:
\begin{align}
\alpha_\mathrm{SG}&=\alpha_\mathrm{GI}\left(\frac{Q_\mathrm{crit}^2}{Q^2}-1\right)\quad\textrm{if $Q<Q_\mathrm{crit}$,}\\
&=0\quad \mathrm{otherwise}.
\end{align}
In the above definitions, we have used the local disk surface density $\Sigma$, its central temperature $T$ and the Toomre parameter $Q=c_s\Omega/(\pi G \Sigma)$ where $c_s$ is the local sound speed, $\Omega$ is the local orbital angular velocity and $G$ is the gravitational constant. In the following, we set $Q_\mathrm{crit}=2$ and $\alpha_\mathrm{GI}=10^{-2}$ while $\alpha_\mathrm{MRI}$, $T_\mathrm{MRI}$ and $\Sigma_\mathrm{crit}$ are allowed to vary.

For the disk cooling, we have two possible opacity tables. In the simplest case, we follow \cite{Martin2011} and assume $\kappa=0.02\,T^{0.8}\,\mathrm{cm}^{2}\,\mathrm{g}^{-1}$. In this case, because the temperature dependence of the opacity is sub-linear, the disk is stable for TI, and we label these models as ``without TI``. In the more elaborated case, we use the full opacity fit of \cite{Armitage2001} (see his Table 1), which include a steep dependence of the opacity when H gets ionised (T>3000\,K), and therefore lead to thermal cycles. These last models are therefore tagged as ``with TI``.

The temperature profile of the MRI+GI instability is shown in Fig \ref{fig:1DMHD_MRI}. The result of the simulation is an effective temperature profile varying in time, which reproduces a radial temperature profile $\propto T_{\text{eff}}\propto r^{-3/4}$, as predicted for a standard accretion disk \citep{SS73}. The instability front extends from the star up to a $\sim 3\mathrm{AU}$ where the MRI is sustained. Locally, in the innermost region up to $\sim 0.3\,\mathrm{AU}$, TI can be sustained and heats up the disk depending of the opacity law chose for this model ('with TI' or 'without TI', see Tab. \ref{tab:simulations}). 
Unlike the TI scenario, starting from a purely theoretical model, MRI+GI enables to obtain solutions commensurable with our observations. The model allows to retrieve the typical size of the outburst, and the compact unresolved emission which naturally arise from the peak of intensity of a power-law temperature profile in the inner region of the disk $<0.1\,\mathrm{AU}$. In the next section, we present in more details the comparison of the model and the data.

\subsection{Parameters of the instability} 
\label{sec:simulations}
We compared the size of the outburst obtained from interferometry with the output of the TI and the MRI-GI models, as shown in Fig. \ref{fig:model-comparison}. 
Following the method of \cite{Malbet2005}, we simulate for each timestep the interferometric visibility and the SED by the numerical integration of successive rings with infinitesimal radius, and an effective temperature given by the output of the MHD simulation. The SED was reddened following the prescription of \cite{Pueyo2012} using $A_v=1.4$, and the lightcurve is extracted from the SEDs computed at each timestep.
Our simulation does not allow to simulate faithfully the inner few stellar radii of the disk, since this would require a dedicated treatment of the boundary layer \citep{Popham1996}, which is out of the scope of this work. Nonetheless, it could be reasonably approximated that the boundary layer follows a power-law temperature profile $T\propto r^{-q}$ in the first $0.1\,\mathrm{AU}$ of FU Orionis, as also visible in \cite{Popham1996}. In order to correct for the inner region, we set the $(q,T_{in},R_{in})$ of the power-law profile of the effective temperature $<0.1\,\mathrm{AU}$ on a joint fit of the SED and the MIRCX data, which accurately probe the inner region of the disk. 
The main parameters which are the least constrained in the MRI-GI scenario are $T_{\mathrm{MRI}}$, $\Sigma_{\mathrm{crit}}$ and $\alpha_\mathrm{MRI}$. In the following, we provide a preliminary exploration of these parameters by comparing the simulations obtained with different values of these parameters. In addition, the runs are separated in pure TI (Simulation 8), pure MRI-GI (Simulations 3-7), and MRI-GI with TI (Simulations 1 \& 2). In order to compare the simulation to our interferometric sizes, we have fitted the same geometrical model to the interferometric visibility from the MHD simulation, adopting the fixed parameters of each respective band as described in Sec.\ref{sec:results}. This method enables to compare equivalent models of the spatial distribution, and to reduce the comparison with the MHD simulation to the evolution of one parameter (fwhm size) in time. 

From the 1D model, we are able to retrieve the typical sizes of the outburst. We reproduce the general morphology of the disk, where the resolved component corresponds mostly to the contribution of the 0.1AU - 1AU inner region, which varies during the outburst, and the unresolved flux, which corresponds to the innermost region $0.1\,\mathrm{AU}$. The typical spatial extent of TI is limited to $0.1\,\mathrm{AU}$, which is too small compared to the measured size. On the contrary, the simulation with MRI-GI with TI seems to provide the best match to the observed sizes and their temporal variation, while MRI-GI without TI seems to produce a disk extension and a temporal variation slightly too large during the outburst.

\begin{figure*}[t!]
\label{fig:SED+lightcurve}
   \centering
   \includegraphics[width=17cm]{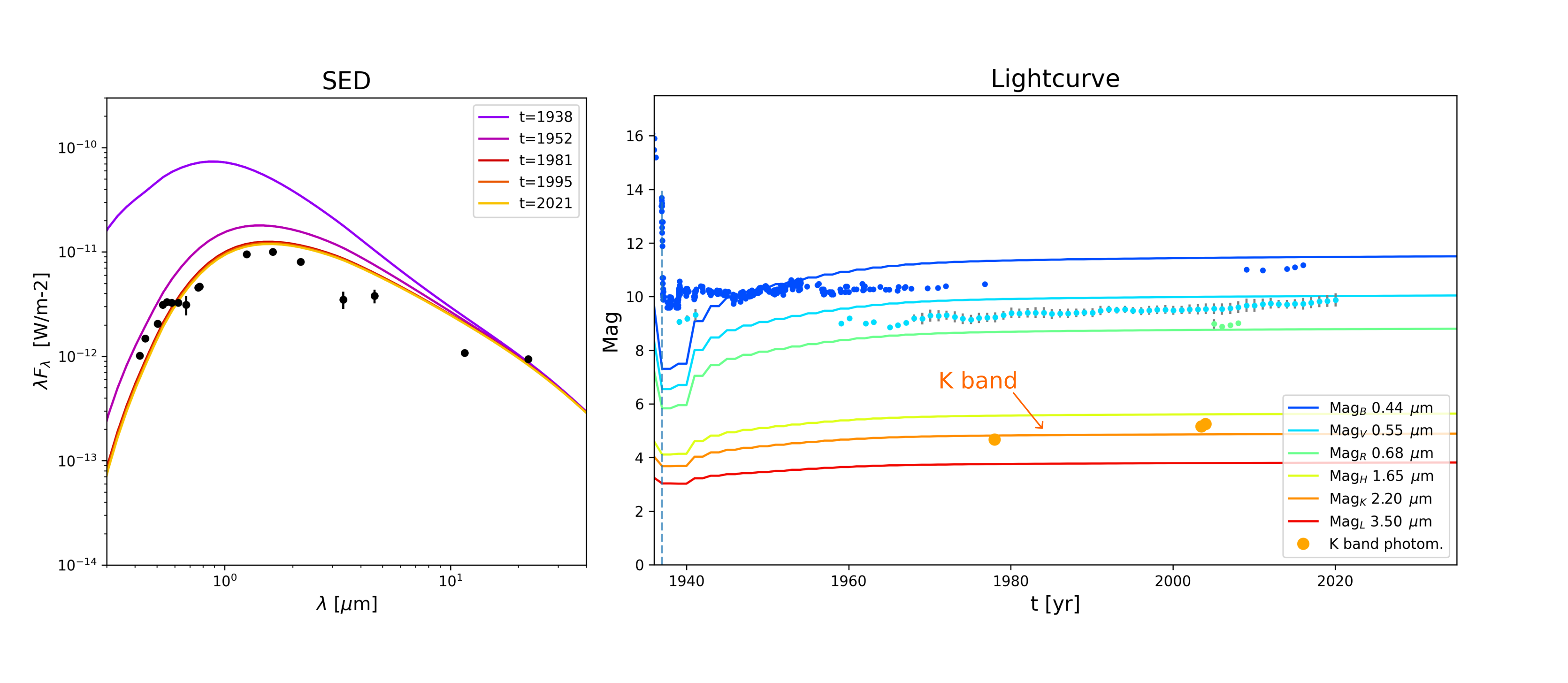}
      \caption{SED (left) and lightcurve (right) simulated from the MHD model and compared to the archival data of FU Orionis.}
\end{figure*}

In the case of the MRI-GI with TI, the parameters which reproduce the instability are $\alpha_\mathrm{MRI}=1.10^{-2}$, $T_{\mathrm{MRI}}=800\,\mathrm{K}$ the activation temperature of the MRI, $\Sigma_{\mathrm{crit}}=10\,\mathrm{g.cm}^{-2}$ the critical density of the disk. A higher value of $\alpha_\mathrm{MRI}=2.10^{-2}$ seems unlikely in this run, as it would translate in a larger size of the outburst (almost a factor 2, Simulation 4). 
Based on the typical size and temporal variation, Simulations 1 \& 2 which includes additional TI on the top of MRI-GI qualitatively better match the typical size of the outburst, however this point is still loosely constrained by the data which are available currently.
For Simulation 1, we obtain $T_{in}=5480\,\mathrm{K}$ and $R_{in}=5.3\,R_{\odot}$, which is consistent with the $R_{in}$ derived in \cite{Malbet2005}. Finally, we can extract the SED and the lightcurve, which are shown in Fig \ref{fig:SED+lightcurve}. The SED and the lightcurves also match the observation data, in particular the K magnitude and its temporal variation. Interestingly, in the case of the lightcurve, the simulation reproduces the small overshoot in B magnitude seen in the visible lightcurve \citep{Herbig1977}. More work is needed to refine the simulated outburst. However, this approach clearly favours the MRI-GI scenario and its basic physics.

\subsection{What triggering event?}
Different events have been proposed to trigger the instability in the outer edge of the outburst region ($\sim 3\,\mathrm{AU}$), so that the instability can propagate outside-in and reproduces the visible lightcurve. The most notable scenario is an instability triggered by an external perturbation, which could be compatible with the presence of a stellar companion FU Ori S at large separation \citep{Bonnell1992, Cuello2020}, or by an internal perturbation, in particular a planet of a few Jupiter mass \citep{Lodato2004}. We note that in the case of MRI-GI, unlike a single fly-by scenario, high-temperature can naturally be sustained in the disk $\sim 5000\,\mathrm{K}$ and with multiple eruptions as seen in some FUors (ZCMa), as long as the disk is replenished in the outer-edge of the dead-zone. MRI-GI scenario does not exclude and could be compatible with an enhancement of GI by an external perturber, which could favor these eruptions in multiple systems still deeply embedded in a reservoir of material. In the specific case of FU Orionis, further investigation on the impact of the tidal interaction of the FU Orionis S (projected separation of 227 AU) on the extended envelope, or planet-signature in the inner region of the disk on the outer edge of the dead-zone, would be important to precise the impact of an external perturbation on the outburst and its dynamic.

\section{Conclusion}
\label{sec:conclusion}
We have presented a temporal monitoring of FU Orionis outburst over 20 years of spatially resolved observations with near-IR interferometry. This dataset was reduced and analyzed homogeneously in order to estimate the size of the outburst region over the time-span covered by interferometric observations. We can draw the following conclusions from this analysis:
\begin{itemize}
    \item the typical structure of FU Orionis can be described by an outbursting region marginally resolved with interferometry, an extended envelope, and a compact flux unresolved in our interferometric data and originating from the inner region of the disk.
    \item we constrain a typical fwhm size of the outbursting region $0.23^{+0.02}_{-0.02}\,\mathrm{AU}$ and $0.28^{+0.02}_{-0.02}\,\mathrm{AU}$ in the H- and K-band respectively, and a temporal variation of $-0.56^{+0.14}_{-0.36}\,\mathrm{AU}/100\,\mathrm{yr}$ and $-0.30^{+0.19}_{-0.19}\,\mathrm{AU}/100\,\mathrm{yr}$ respectively. This size variation is compatible with a constant or a slowly decreasing size of the outbursting region.
    \item the most consistent model is a scenario in which MRI is triggered by a GI instability at the outer edge zone, occurring in a magnetically layered disk, and reproduces consistently the interferometric sizes and visibilities, the SED and the lightcurve observed in FU Orionis.
    \item these preliminary 1D simulations are mostly compatible with $\alpha_\mathrm{MRI}=1.10^{-2}$, $T_{\mathrm{MRI}}=800\,K$ and $\Sigma_{\mathrm{crit}}=10\,g.cm^{-2}$. Higher values of $\alpha_{\mathrm{MRI}}$ are less favoured as they produce larger outburst regions. Further work is needed to confirm the value of these parameters both on observations and models.
    \item this scenario could be compatible with an external perturber enhancing GI at the outer-edge of the dead-zone e.g. through tidal interaction with a large separation stellar companion or a planet at the outer edge of the dead-zone.
\end{itemize}
Our study refines the scenario of FU Orionis outburst. This highlights the unique potential of directly probing the inner AU of protoplanetary disks with spatially resolved observations to understand the process of planet formation.

\begin{acknowledgements}
G.B and J-P.B thank Regis Lachaume for having provided the PTI, IOTA and VINCI archival data, and Pierre Kervella for providing the data reduction software of VINCI. We would like to thank the individuals who have built and contributed to develop during more than 20 years the generations of interferometric instruments whose data are included here. This research has benefited from the support of the Jean-Marie Mariotti Center (JMMC), and has made use of \texttt{Aspro} and \texttt{SearchCal} services \footnote{Available at http://www.jmmc.fr}.
\end{acknowledgements}

%
%
\bibliographystyle{aa}
\bibliography{PaperBib} 

\begin{thebibliography}{42}
\expandafter\ifx\csname natexlab\endcsname\relax\def\natexlab#1{#1}\fi

\bibitem[{{Armitage}(2011)}]{Armitage2011}
{Armitage}, P.~J. 2011, \araa, 49, 195

\bibitem[{{Armitage} {et~al.}(2001){Armitage}, {Livio}, \&
  {Pringle}}]{Armitage2001}
{Armitage}, P.~J., {Livio}, M., \& {Pringle}, J.~E. 2001, \mnras, 324, 705

\bibitem[{{Audard} {et~al.}(2014){Audard}, {{\'A}brah{\'a}m}, {Dunham},
  {Green}, {Grosso}, {Hamaguchi}, {Kastner}, {K{\'o}sp{\'a}l}, {Lodato},
  {Romanova}, {Skinner}, {Vorobyov}, \& {Zhu}}]{Audard2014}
{Audard}, M., {{\'A}brah{\'a}m}, P., {Dunham}, M.~M., {et~al.} 2014, in
  Protostars and Planets VI, ed. H.~{Beuther}, R.~S. {Klessen}, C.~P.
  {Dullemond}, \& T.~{Henning}, 387

\bibitem[{{Balbus} \& {Hawley}(1998)}]{Balbus1998}
{Balbus}, S.~A. \& {Hawley}, J.~F. 1998, Reviews of Modern Physics, 70, 1

\bibitem[{{Bell} \& {Lin}(1994)}]{Bell1994}
{Bell}, K.~R. \& {Lin}, D.~N.~C. 1994, \apj, 427, 987

\bibitem[{{Bonnell} \& {Bastien}(1992)}]{Bonnell1992}
{Bonnell}, I. \& {Bastien}, P. 1992, \apjl, 401, L31

\bibitem[{{Calvet} {et~al.}(2000){Calvet}, {Hartmann}, \& {Strom}}]{Calvet2000}
{Calvet}, N., {Hartmann}, L., \& {Strom}, S.~E. 2000, in Protostars and Planets
  IV, ed. V.~{Mannings}, A.~P. {Boss}, \& S.~S. {Russell}, 377

\bibitem[{{Cuello} {et~al.}(2020){Cuello}, {Louvet}, {Mentiplay}, {Pinte},
  {Price}, {Winter}, {Nealon}, {M{\'e}nard}, {Lodato}, {Dipierro},
  {Christiaens}, {Montesinos}, {Cuadra}, {Laibe}, {Cieza}, {Dong}, \&
  {Alexander}}]{Cuello2020}
{Cuello}, N., {Louvet}, F., {Mentiplay}, D., {et~al.} 2020, \mnras, 491, 504

\bibitem[{{Dong} {et~al.}(2016){Dong}, {Vorobyov}, {Pavlyuchenkov}, {Chiang},
  \& {Liu}}]{Dong2016}
{Dong}, R., {Vorobyov}, E., {Pavlyuchenkov}, Y., {Chiang}, E., \& {Liu}, H.~B.
  2016, \apj, 823, 141

\bibitem[{{Efron}(1982)}]{Efron1982}
{Efron}, B. 1982, {The Jackknife, the Bootstrap and other resampling plans}

\bibitem[{{Gaia Collaboration} {et~al.}(2021){Gaia Collaboration}, {Brown},
  {Vallenari}, {Prusti}, {de Bruijne}, {Babusiaux}, {Biermann}, {Creevey},
  {Evans}, {Eyer}, {Hutton}, {Jansen}, {Jordi}, {Klioner}, {Lammers},
  {Lindegren}, {Luri}, {Mignard}, {Panem}, {Pourbaix}, {Randich}, {Sartoretti},
  {Soubiran}, {Walton}, {Arenou}, {Bailer-Jones}, {Bastian}, {Cropper},
  {Drimmel}, {Katz}, {Lattanzi}, {van Leeuwen}, {Bakker}, {Cacciari},
  {Casta{\~n}eda}, {De Angeli}, {Ducourant}, {Fabricius}, {Fouesneau},
  {Fr{\'e}mat}, {Guerra}, {Guerrier}, {Guiraud}, {Jean-Antoine Piccolo},
  {Masana}, {Messineo}, {Mowlavi}, {Nicolas}, {Nienartowicz}, {Pailler},
  {Panuzzo}, {Riclet}, {Roux}, {Seabroke}, {Sordo}, {Tanga}, {Th{\'e}venin},
  {Gracia-Abril}, {Portell}, {Teyssier}, {Altmann}, {Andrae}, {Bellas-Velidis},
  {Benson}, {Berthier}, {Blomme}, {Brugaletta}, {Burgess}, {Busso}, {Carry},
  {Cellino}, {Cheek}, {Clementini}, {Damerdji}, {Davidson}, {Delchambre},
  {Dell'Oro}, {Fern{\'a}ndez-Hern{\'a}ndez}, {Galluccio}, {Garc{\'\i}a-Lario},
  {Garcia-Reinaldos}, {Gonz{\'a}lez-N{\'u}{\~n}ez}, {Gosset}, {Haigron},
  {Halbwachs}, {Hambly}, {Harrison}, {Hatzidimitriou}, {Heiter},
  {Hern{\'a}ndez}, {Hestroffer}, {Hodgkin}, {Holl}, {Jan{\ss}en}, {Jevardat de
  Fombelle}, {Jordan}, {Krone-Martins}, {Lanzafame}, {L{\"o}ffler}, {Lorca},
  {Manteiga}, {Marchal}, {Marrese}, {Moitinho}, {Mora}, {Muinonen}, {Osborne},
  {Pancino}, {Pauwels}, {Petit}, {Recio-Blanco}, {Richards}, {Riello},
  {Rimoldini}, {Robin}, {Roegiers}, {Rybizki}, {Sarro}, {Siopis}, {Smith},
  {Sozzetti}, {Ulla}, {Utrilla}, {van Leeuwen}, {van Reeven}, {Abbas}, {Abreu
  Aramburu}, {Accart}, {Aerts}, {Aguado}, {Ajaj}, {Altavilla}, {{\'A}lvarez},
  {{\'A}lvarez Cid-Fuentes}, {Alves}, {Anderson}, {Anglada Varela}, {Antoja},
  {Audard}, {Baines}, {Baker}, {Balaguer-N{\'u}{\~n}ez}, {Balbinot}, {Balog},
  {Barache}, {Barbato}, {Barros}, {Barstow}, {Bartolom{\'e}}, {Bassilana},
  {Bauchet}, {Baudesson-Stella}, {Becciani}, {Bellazzini}, {Bernet}, {Bertone},
  {Bianchi}, {Blanco-Cuaresma}, {Boch}, {Bombrun}, {Bossini}, {Bouquillon},
  {Bragaglia}, {Bramante}, {Breedt}, {Bressan}, {Brouillet}, {Bucciarelli},
  {Burlacu}, {Busonero}, {Butkevich}, {Buzzi}, {Caffau}, {Cancelliere},
  {C{\'a}novas}, {Cantat-Gaudin}, {Carballo}, {Carlucci}, {Carnerero},
  {Carrasco}, {Casamiquela}, {Castellani}, {Castro-Ginard}, {Castro Sampol},
  {Chaoul}, {Charlot}, {Chemin}, {Chiavassa}, {Cioni}, {Comoretto}, {Cooper},
  {Cornez}, {Cowell}, {Crifo}, {Crosta}, {Crowley}, {Dafonte}, {Dapergolas},
  {David}, {David}, {de Laverny}, {De Luise}, {De March}, {De Ridder}, {de
  Souza}, {de Teodoro}, {de Torres}, {del Peloso}, {del Pozo}, {Delbo},
  {Delgado}, {Delgado}, {Delisle}, {Di Matteo}, {Diakite}, {Diener},
  {Distefano}, {Dolding}, {Eappachen}, {Edvardsson}, {Enke}, {Esquej}, {Fabre},
  {Fabrizio}, {Faigler}, {Fedorets}, {Fernique}, {Fienga}, {Figueras},
  {Fouron}, {Fragkoudi}, {Fraile}, {Franke}, {Gai}, {Garabato},
  {Garcia-Gutierrez}, {Garc{\'\i}a-Torres}, {Garofalo}, {Gavras}, {Gerlach},
  {Geyer}, {Giacobbe}, {Gilmore}, {Girona}, {Giuffrida}, {Gomel}, {Gomez},
  {Gonzalez-Santamaria}, {Gonz{\'a}lez-Vidal}, {Granvik},
  {Guti{\'e}rrez-S{\'a}nchez}, {Guy}, {Hauser}, {Haywood}, {Helmi}, {Hidalgo},
  {Hilger}, {H{\l}adczuk}, {Hobbs}, {Holland}, {Huckle}, {Jasniewicz},
  {Jonker}, {Juaristi Campillo}, {Julbe}, {Karbevska}, {Kervella}, {Khanna},
  {Kochoska}, {Kontizas}, {Kordopatis}, {Korn}, {Kostrzewa-Rutkowska},
  {Kruszy{\'n}ska}, {Lambert}, {Lanza}, {Lasne}, {Le Campion}, {Le Fustec},
  {Lebreton}, {Lebzelter}, {Leccia}, {Leclerc}, {Lecoeur-Taibi}, {Liao},
  {Licata}, {Lindstr{\o}m}, {Lister}, {Livanou}, {Lobel}, {Madrero Pardo},
  {Managau}, {Mann}, {Marchant}, {Marconi}, {Marcos Santos}, {Marinoni},
  {Marocco}, {Marshall}, {Martin Polo}, {Mart{\'\i}n-Fleitas}, {Masip},
  {Massari}, {Mastrobuono-Battisti}, {Mazeh}, {McMillan}, {Messina},
  {Michalik}, {Millar}, {Mints}, {Molina}, {Molinaro}, {Moln{\'a}r},
  {Montegriffo}, {Mor}, {Morbidelli}, {Morel}, {Morris}, {Mulone}, {Munoz},
  {Muraveva}, {Murphy}, {Musella}, {Noval}, {Ord{\'e}novic}, {Orr{\`u}},
  {Osinde}, {Pagani}, {Pagano}, {Palaversa}, {Palicio}, {Panahi}, {Pawlak},
  {Pe{\~n}alosa Esteller}, {Penttil{\"a}}, {Piersimoni}, {Pineau}, {Plachy},
  {Plum}, {Poggio}, {Poretti}, {Poujoulet}, {Pr{\v{s}}a}, {Pulone}, {Racero},
  {Ragaini}, {Rainer}, {Raiteri}, {Rambaux}, {Ramos}, {Ramos-Lerate}, {Re
  Fiorentin}, {Regibo}, {Reyl{\'e}}, {Ripepi}, {Riva}, {Rixon}, {Robichon},
  {Robin}, {Roelens}, {Rohrbasser}, {Romero-G{\'o}mez}, {Rowell}, {Royer},
  {Rybicki}, {Sadowski}, {Sagrist{\`a} Sell{\'e}s}, {Sahlmann}, {Salgado},
  {Salguero}, {Samaras}, {Sanchez Gimenez}, {Sanna}, {Santove{\~n}a},
  {Sarasso}, {Schultheis}, {Sciacca}, {Segol}, {Segovia}, {S{\'e}gransan},
  {Semeux}, {Shahaf}, {Siddiqui}, {Siebert}, {Siltala}, {Slezak}, {Smart},
  {Solano}, {Solitro}, {Souami}, {Souchay}, {Spagna}, {Spoto}, {Steele},
  {Steidelm{\"u}ller}, {Stephenson}, {S{\"u}veges}, {Szabados}, {Szegedi-Elek},
  {Taris}, {Tauran}, {Taylor}, {Teixeira}, {Thuillot}, {Tonello}, {Torra},
  {Torra}, {Turon}, {Unger}, {Vaillant}, {van Dillen}, {Vanel}, {Vecchiato},
  {Viala}, {Vicente}, {Voutsinas}, {Weiler}, {Wevers}, {Wyrzykowski}, {Yoldas},
  {Yvard}, {Zhao}, {Zorec}, {Zucker}, {Zurbach}, \& {Zwitter}}]{GaiaEDR3}
{Gaia Collaboration}, {Brown}, A.~G.~A., {Vallenari}, A., {et~al.} 2021, \aap,
  649, A1

\bibitem[{{Gammie}(1996)}]{Gammie1996}
{Gammie}, C.~F. 1996, \apj, 457, 355

\bibitem[{{Gravity Collaboration} {et~al.}(2019){Gravity Collaboration},
  {Perraut}, {Labadie}, {Lazareff}, {Klarmann}, {Segura-Cox}, {Benisty},
  {Bouvier}, {Brandner}, {Caratti O Garatti}, {Caselli}, {Dougados}, {Garcia},
  {Garcia-Lopez}, {Kendrew}, {Koutoulaki}, {Kervella}, {Lin}, {Pineda},
  {Sanchez-Bermudez}, {van Dishoeck}, {Abuter}, {Amorim}, {Berger}, {Bonnet},
  {Buron}, {Cantalloube}, {Cl{\'e}net}, {Coud{\'e} Du Foresto}, {Dexter}, {de
  Zeeuw}, {Duvert}, {Eckart}, {Eisenhauer}, {Eupen}, {Gao}, {Gendron},
  {Genzel}, {Gillessen}, {Gordo}, {Grellmann}, {Haubois}, {Haussmann},
  {Henning}, {Hippler}, {Horrobin}, {Hubert}, {Jocou}, {Lacour}, {Le Bouquin},
  {L{\'e}na}, {M{\'e}rand}, {Ott}, {Paumard}, {Perrin}, {Pfuhl}, {Rabien},
  {Ray}, {Rau}, {Rousset}, {Scheithauer}, {Straub}, {Straubmeier}, {Sturm},
  {Vincent}, {Waisberg}, {Wank}, {Widmann}, {Wieprecht}, {Wiest}, {Wiezorrek},
  {Woillez}, \& {Yazici}}]{Perraut2019}
{Gravity Collaboration}, {Perraut}, K., {Labadie}, L., {et~al.} 2019, \aap,
  632, A53

\bibitem[{{Hartmann} \& {Kenyon}(1996)}]{Hartmann1996}
{Hartmann}, L. \& {Kenyon}, S.~J. 1996, \araa, 34, 207

\bibitem[{{Herbig}(1977)}]{Herbig1977}
{Herbig}, G.~H. 1977, \apj, 217, 693

\bibitem[{{Hirose}(2015)}]{Hirose2015}
{Hirose}, S. 2015, \mnras, 448, 3105

\bibitem[{{Kenyon} \& {Hartmann}(1990)}]{Kenyon1990}
{Kenyon}, S.~J. \& {Hartmann}, L.~W. 1990, \apj, 349, 197

\bibitem[{{Kervella} {et~al.}(2004){Kervella}, {S{\'e}gransan}, \& {Coud{\'e}
  du Foresto}}]{Kervella2004}
{Kervella}, P., {S{\'e}gransan}, D., \& {Coud{\'e} du Foresto}, V. 2004, \aap,
  425, 1161

\bibitem[{{Labdon} {et~al.}(2020){Labdon}, {Kraus}, {Davies}, {Kreplin},
  {Monnier}, {Le Bouquin}, {Anugu}, {Brummelaar}, {Setterholm}, {Gardener},
  {Ennis}, {Lanthermann}, {Schaefer}, \& {Laws}}]{Labdon2020}
{Labdon}, A., {Kraus}, S., {Davies}, C.~L., {et~al.} 2020, arXiv e-prints,
  arXiv:2011.07865

\bibitem[{{Lachaume} {et~al.}(2019){Lachaume}, {Rabus}, {Jord{\'a}n}, {Brahm},
  {Boyajian}, {von Braun}, \& {Berger}}]{Lachaume2019}
{Lachaume}, R., {Rabus}, M., {Jord{\'a}n}, A., {et~al.} 2019, \mnras, 484, 2656

\bibitem[{{Lazareff} {et~al.}(2017){Lazareff}, {Berger}, {Kluska}, {Le
  Bouquin}, {Benisty}, {Malbet}, {Koen}, {Pinte}, {Thi}, {Absil}, {Baron},
  {Delboulb{\'e}}, {Duvert}, {Isella}, {Jocou}, {Juhasz}, {Kraus}, {Lachaume},
  {M{\'e}nard}, {Millan-Gabet}, {Monnier}, {Moulin}, {Perraut}, {Rochat},
  {Soulez}, {Tallon}, {Thi{\'e}baut}, {Traub}, \& {Zins}}]{Lazareff2017}
{Lazareff}, B., {Berger}, J.~P., {Kluska}, J., {et~al.} 2017, \aap, 599, A85

\bibitem[{{Le Bouquin} {et~al.}(2011){Le Bouquin}, {Berger}, {Lazareff},
  {Zins}, {Haguenauer}, {Jocou}, {Kern}, {Millan-Gabet}, {Traub}, {Absil},
  {Augereau}, {Benisty}, {Blind}, {Bonfils}, {Bourget}, {Delboulbe},
  {Feautrier}, {Germain}, {Gitton}, {Gillier}, {Kiekebusch}, {Kluska},
  {Knudstrup}, {Labeye}, {Lizon}, {Monin}, {Magnard}, {Malbet}, {Maurel},
  {M{\'e}nard}, {Micallef}, {Michaud}, {Montagnier}, {Morel}, {Moulin},
  {Perraut}, {Popovic}, {Rabou}, {Rochat}, {Rojas}, {Roussel}, {Roux},
  {Stadler}, {Stefl}, {Tatulli}, \& {Ventura}}]{LeBouquin2011}
{Le Bouquin}, J.~B., {Berger}, J.~P., {Lazareff}, B., {et~al.} 2011, \aap, 535,
  A67

\bibitem[{{Lesur} {et~al.}(2022){Lesur}, {Ercolano}, {Flock}, {Lin}, {Yang},
  {Barranco}, {Benitez-Llambay}, {Goodman}, {Johansen}, {Klahr}, {Laibe},
  {Lyra}, {Marcus}, {Nelson}, {Squire}, {Simon}, {Turner}, {Umurhan}, \&
  {Youdin}}]{Lesur22}
{Lesur}, G., {Ercolano}, B., {Flock}, M., {et~al.} 2022, arXiv e-prints,
  arXiv:2203.09821

\bibitem[{{Liu} {et~al.}(2019){Liu}, {M{\'e}rand}, {Green}, {P{\'e}rez},
  {Hales}, {Yang}, {Dunham}, {Hasegawa}, {Henning}, {Galv{\'a}n-Madrid},
  {K{\'o}sp{\'a}l}, {Takami}, {Vorobyov}, \& {Zhu}}]{Liu2019}
{Liu}, H.~B., {M{\'e}rand}, A., {Green}, J.~D., {et~al.} 2019, \apj, 884, 97

\bibitem[{{Lodato} \& {Clarke}(2004)}]{Lodato2004}
{Lodato}, G. \& {Clarke}, C.~J. 2004, \mnras, 353, 841

\bibitem[{{Lykou} {et~al.}(2022){Lykou}, {{\'A}brah{\'a}m}, {Chen}, {Varga},
  {K{\'o}sp{\'a}l}, {Matter}, {Siwak}, {Szab{\'o}}, {Zhu}, {Liu}, {Lopez},
  {Allouche}, {Augereau}, {Berio}, {Cruzal{\`e}bes}, {Dominik}, {Henning},
  {Hofmann}, {Hogerheijde}, {Jaffe}, {Kokoulina}, {Lagarde}, {Meilland},
  {Millour}, {Pantin}, {Petrov}, {Robbe-Dubois}, {Schertl}, {Scheuck}, {van
  Boekel}, {Waters}, {Weigelt}, \& {Wolf}}]{MATISSE2022}
{Lykou}, F., {{\'A}brah{\'a}m}, P., {Chen}, L., {et~al.} 2022, \aap, 663, A86

\bibitem[{{Malbet} {et~al.}(1998){Malbet}, {Berger}, {Colavita}, {Koresko},
  {Beichman}, {Boden}, {Kulkarni}, {Lane}, {Mobley}, {Pan}, {Shao}, {Van
  Belle}, \& {Wallace}}]{Malbet1998}
{Malbet}, F., {Berger}, J.~P., {Colavita}, M.~M., {et~al.} 1998, \apjl, 507,
  L149

\bibitem[{{Malbet} {et~al.}(2005){Malbet}, {Lachaume}, {Berger}, {Colavita},
  {di Folco}, {Eisner}, {Lane}, {Millan-Gabet}, {S{\'e}gransan}, \&
  {Traub}}]{Malbet2005}
{Malbet}, F., {Lachaume}, R., {Berger}, J.~P., {et~al.} 2005, \aap, 437, 627

\bibitem[{{Martin} \& {Lubow}(2011)}]{Martin2011}
{Martin}, R.~G. \& {Lubow}, S.~H. 2011, \apjl, 740, L6

\bibitem[{{Millan-Gabet} {et~al.}(2006){Millan-Gabet}, {Monnier}, {Akeson},
  {Hartmann}, {Berger}, {Tannirkulam}, {Melnikov}, {Billmeier}, {Calvet},
  {D'Alessio}, {Hillenbrand}, {Kuchner}, {Traub}, {Tuthill}, {Beichman},
  {Boden}, {Booth}, {Colavita}, {Creech-Eakman}, {Gathright}, {Hrynevych},
  {Koresko}, {Le Mignant}, {Ligon}, {Mennesson}, {Neyman}, {Sargent}, {Shao},
  {Swain}, {Thompson}, {Unwin}, {van Belle}, {Vasisht}, \&
  {Wizinowich}}]{Millan-Gabet2006}
{Millan-Gabet}, R., {Monnier}, J.~D., {Akeson}, R.~L., {et~al.} 2006, \apj,
  641, 547

\bibitem[{{P{\'e}rez} {et~al.}(2020){P{\'e}rez}, {Hales}, {Liu}, {Zhu},
  {Casassus}, {Williams}, {Zurlo}, {Cuello}, {Cieza}, \&
  {Principe}}]{Perez2020}
{P{\'e}rez}, S., {Hales}, A., {Liu}, H.~B., {et~al.} 2020, \apj, 889, 59

\bibitem[{{Popham} {et~al.}(1996){Popham}, {Kenyon}, {Hartmann}, \&
  {Narayan}}]{Popham1996}
{Popham}, R., {Kenyon}, S., {Hartmann}, L., \& {Narayan}, R. 1996, \apj, 473,
  422

\bibitem[{{Pueyo} {et~al.}(2012){Pueyo}, {Hillenbrand}, {Vasisht},
  {Oppenheimer}, {Monnier}, {Hinkley}, {Crepp}, {Roberts}, {Brenner},
  {Zimmerman}, {Parry}, {Beichman}, {Dekany}, {Shao}, {Burruss}, {Cady},
  {Roberts}, \& {Soummer}}]{Pueyo2012}
{Pueyo}, L., {Hillenbrand}, L., {Vasisht}, G., {et~al.} 2012, \apj, 757, 57

\bibitem[{{Quanz} {et~al.}(2006){Quanz}, {Henning}, {Bouwman}, {Ratzka}, \&
  {Leinert}}]{Quanz2006}
{Quanz}, S.~P., {Henning}, T., {Bouwman}, J., {Ratzka}, T., \& {Leinert}, C.
  2006, \apj, 648, 472

\bibitem[{{Scepi} {et~al.}(2019){Scepi}, {Dubus}, \& {Lesur}}]{Scepi19}
{Scepi}, N., {Dubus}, G., \& {Lesur}, G. 2019, \aap, 626, A116

\bibitem[{{Scepi} {et~al.}(2020){Scepi}, {Lesur}, {Dubus}, \&
  {Jacquemin-Ide}}]{Scepi20}
{Scepi}, N., {Lesur}, G., {Dubus}, G., \& {Jacquemin-Ide}, J. 2020, \aap, 641,
  A133

\bibitem[{{Shakura} \& {Sunyaev}(1973)}]{SS73}
{Shakura}, N.~I. \& {Sunyaev}, R.~A. 1973, \aap, 24, 337

\bibitem[{{Takami} {et~al.}(2018){Takami}, {Fu}, {Liu}, {Karr}, {Hashimoto},
  {Kudo}, {Vorobyov}, {K{\'o}sp{\'a}l}, {Scicluna}, {Dong}, {Tamura}, {Pyo},
  {Fukagawa}, {Tsuribe}, {Dunham}, {Henning}, \& {de Leon}}]{Takami2018}
{Takami}, M., {Fu}, G., {Liu}, H.~B., {et~al.} 2018, \apj, 864, 20

\bibitem[{{Toomre}(1964)}]{Toomre64}
{Toomre}, A. 1964, \apj, 139, 1217

\bibitem[{{Vorobyov} \& {Basu}(2005)}]{Vorobyov2005}
{Vorobyov}, E.~I. \& {Basu}, S. 2005, \apjl, 633, L137

\bibitem[{{Zhu} {et~al.}(2007){Zhu}, {Hartmann}, {Calvet}, {Hernandez},
  {Muzerolle}, \& {Tannirkulam}}]{Zhu2007}
{Zhu}, Z., {Hartmann}, L., {Calvet}, N., {et~al.} 2007, \apj, 669, 483

\bibitem[{{Zhu} {et~al.}(2009){Zhu}, {Hartmann}, {Gammie}, \&
  {McKinney}}]{Zhu2009}
{Zhu}, Z., {Hartmann}, L., {Gammie}, C., \& {McKinney}, J.~C. 2009, \apj, 701,
  620

\end{thebibliography}

\begin{appendix} 

\onecolumn
\section{Output of the bootstrap analysis in H and K band}

\begin{figure*}[!htb]
   \centering
   \includegraphics[width=12.0cm]{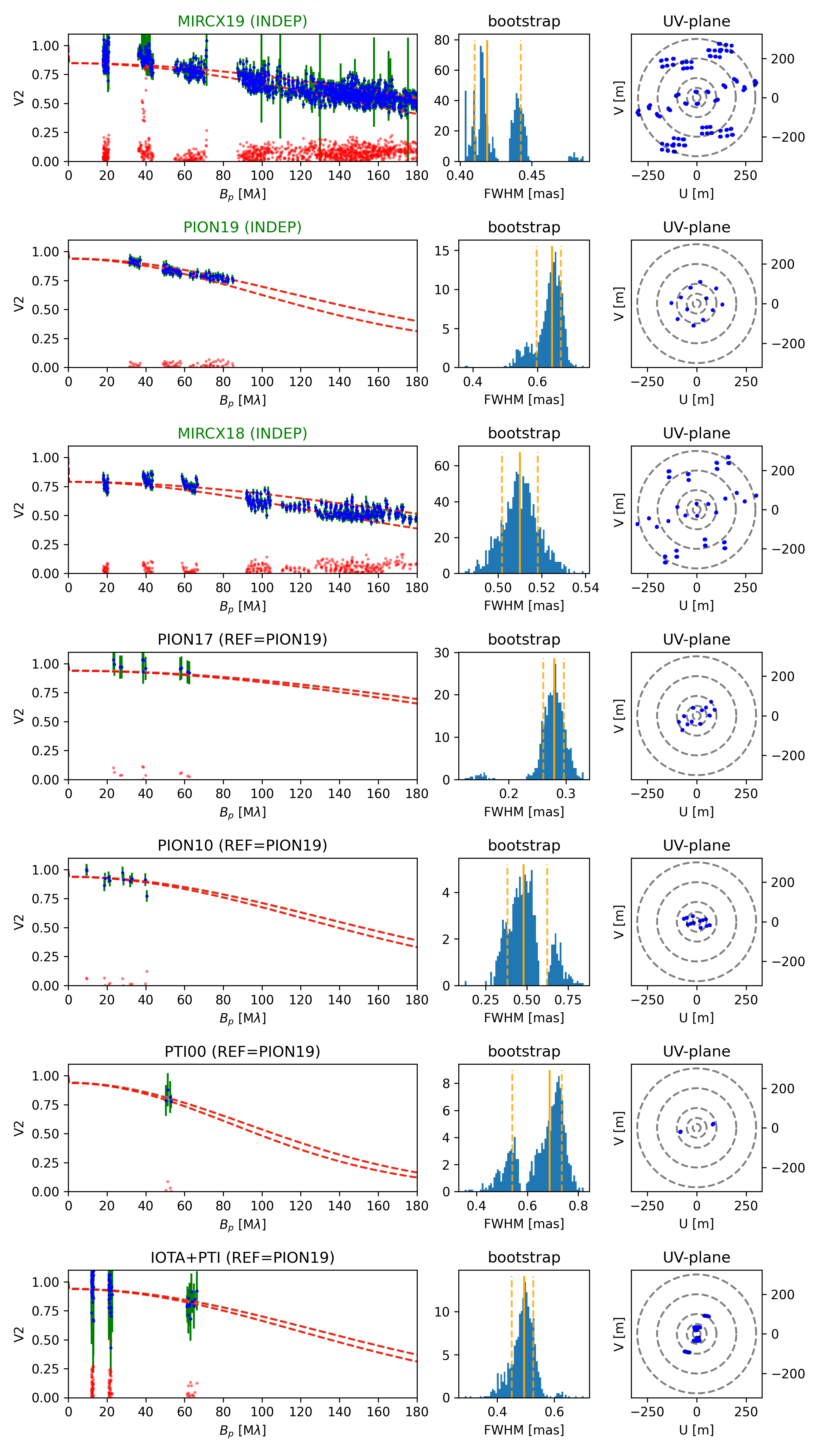}
      \caption{H band : output of the fitting procedure of the interferometric observations. \textbf{Left :} The visibility model is represented in dashed line, the minor axis of the gaussian disk is associated to the upper visibility curve and the major axis to the lower visibility curve. Residuals are shown in red scattered points. \textbf{Middle :} Bootstrap values of the fwhm estimate. \textbf{Right :} (u,v)-coverage associated to each observation.}
         \label{fig:outH}
\end{figure*}

\newpage
\begin{figure*}[!htb]
   \centering
   \includegraphics[width=13.0cm]{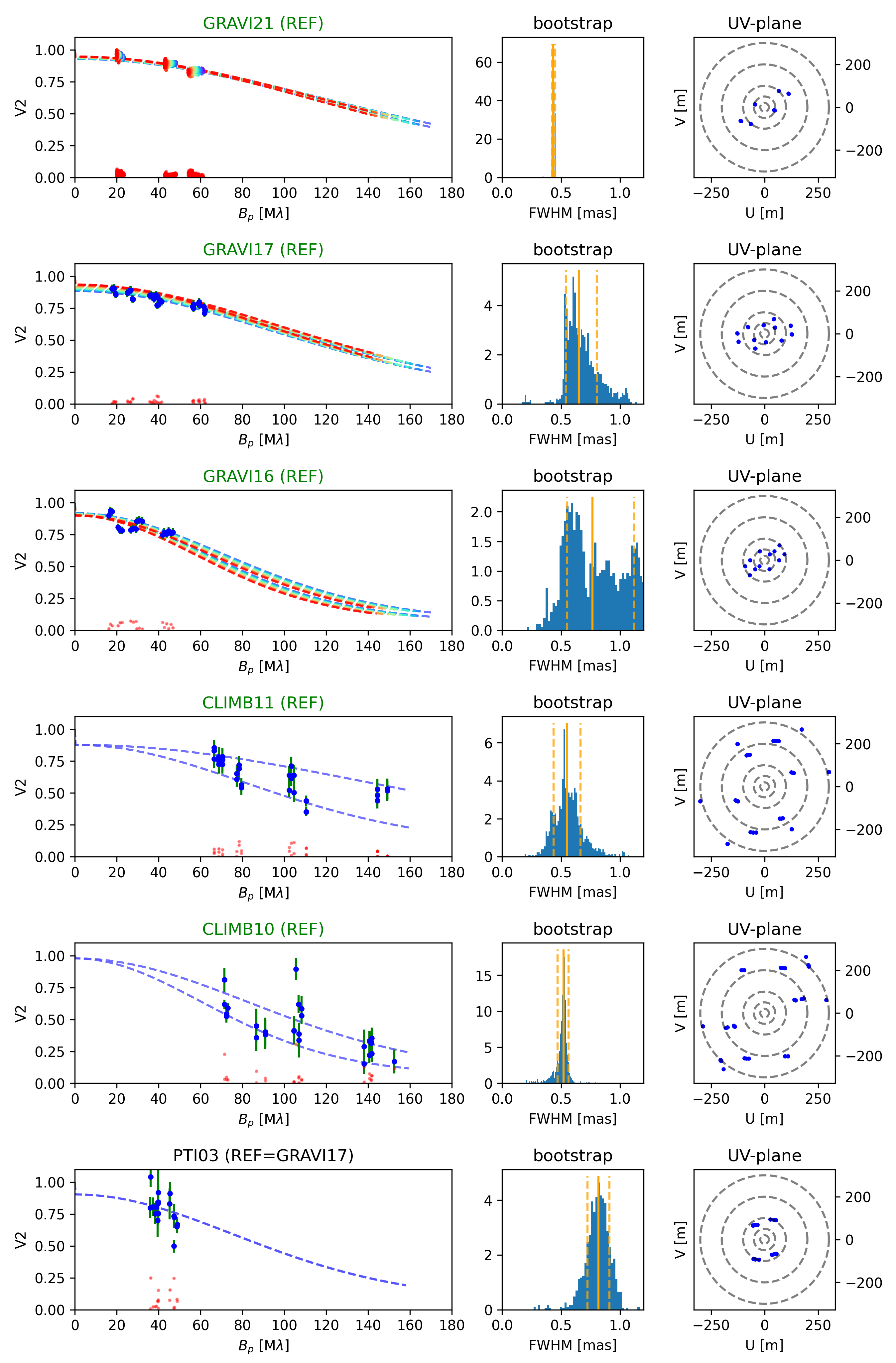}
      \caption{K band : output of the fitting procedure. See Fig.\ref{fig:outH} for details of the plot, and Fig.\ref{fig:outK_pII} for the second part.
              }
         \label{fig:outK_pI}
\end{figure*}

\newpage
\begin{figure*}[!htb]
   \centering
   \includegraphics[width=13.0cm]{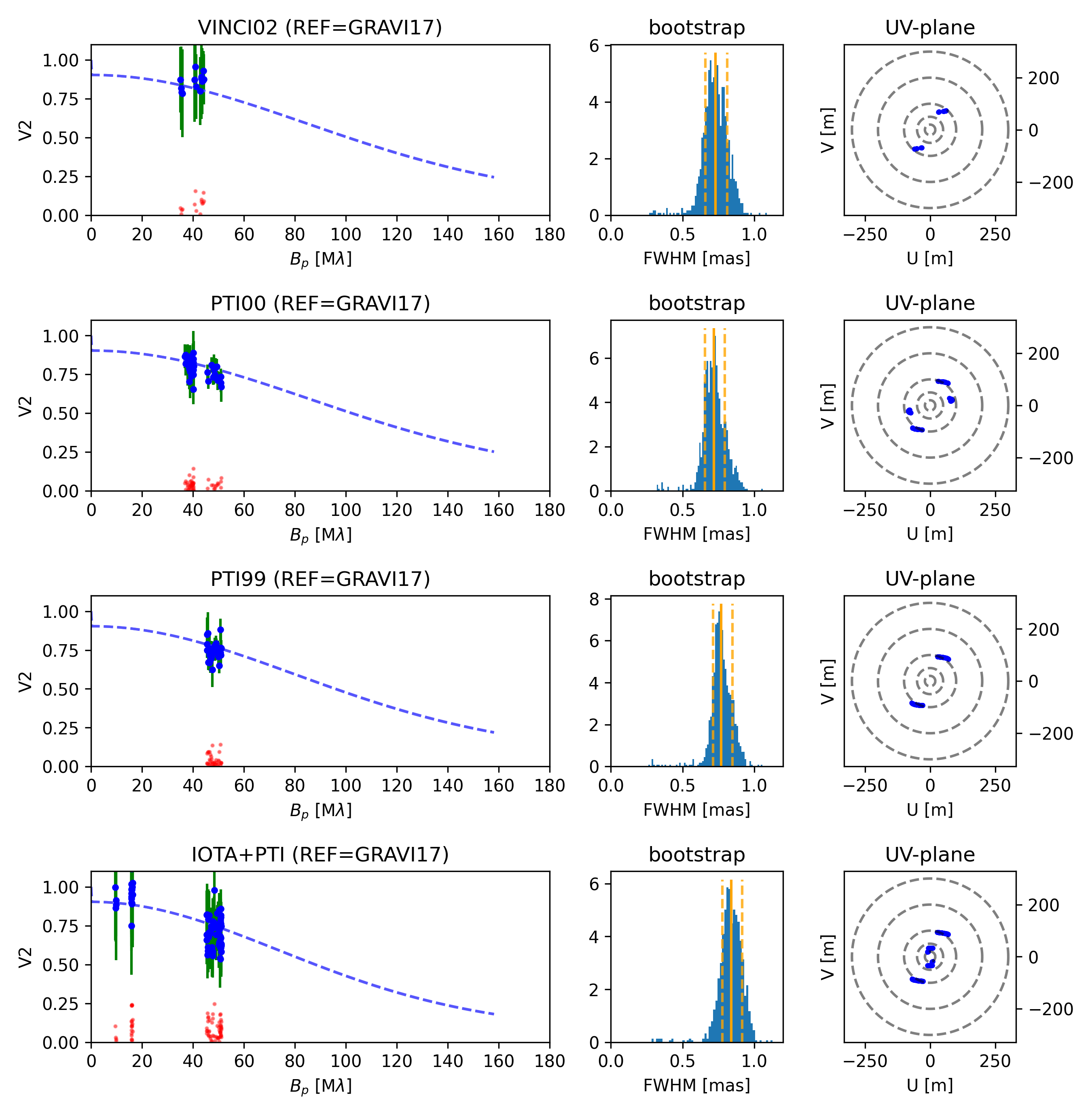}
   \renewcommand{\figurename}{\textbf{Fig. A.2 Continued}}
   \renewcommand{\thefigure}{}
   \caption{}
   \label{fig:outK_pII}
\end{figure*}

\newpage

\section{Fitted parameters}

\begin{table*}[!htb]
\centering                          
\renewcommand{\arraystretch}{1.2}
\resizebox{\textwidth}{!}{
\begin{tabular}{c c c c c c c c c c} 
\hline\hline                
& & & \multicolumn{7}{c}{Fitted parameters} \\ \cline{4-10} 
H Band & & & $a$ & $\theta$ & $i$ & $f_{c}$  & $f_0$ & $f_1$ & $f_2$\\  
 & &  Obs & [mas] & [$^\circ$] & [$^\circ$] & [\%] & [\%] & $[\%]/\mu\mathrm{m}$ & $[\%]/\mu\mathrm{m}^2$\\ \cline{2-10}   
& (1) & MIRCX19 & $0.46^{+0.01}_{-0.01}$& $47^{+2}_{-2}$& $44^{+1}_{-1}$& $27.7^{+0.3}_{-0.3}$& $8.5^{+0.1}_{-0.1}$& $-0.0^{+0.1}_{-0.1}$& $0.0^{+0.0}_{-0.0}$\\
& \textbf{(2)} & \textbf{PION19} & $\mathbf{0.62^{+0.03}_{-0.03}}$& $\mathbf{51^{+6}_{-5}}$& $\mathbf{32^{+4}_{-4}}$& $\mathbf{16.7}$& $\mathbf{2.8^{+1.0}_{-1.0}}$& $\mathbf{1.5^{+2.0}_{-2.0}}$& $\mathbf{0.0^{+0.0}_{-0.0}}$\\
& (3) & MIRCX18 & $0.49^{+0.01}_{-0.01}$& $56^{+2}_{-2}$& $41^{+1}_{-1}$& $29.2^{+0.7}_{-1.0}$& $10.2^{+0.2}_{-0.2}$& $0.0^{+0.1}_{-0.1}$& $0.0^{+0.0}_{-0.0}$\\
& (4) & PION17 & $0.89^{+0.45}_{-0.21}$& $56^{+14}_{-14}$& $89^{+30}_{-30}$& $\mathtt{16.7}$& $3.0^{+1.7}_{-1.7}$& $-33.4^{+12.8}_{-12.8}$& $0.0^{+0.0}_{-0.0}$\\
& (5) & PION10 & $0.63^{+0.22}_{-0.17}$& $\mathtt{51}$& $\mathtt{32}$& $\mathtt{29.2}$& $\mathtt{2.8}$& $\mathtt{1.5}$& $\mathtt{0.0}$\\
& (6) & PTI00 & $0.82^{+0.10}_{-0.05}$& $\mathtt{51}$& $\mathtt{32}$& $\mathtt{29.2}$& $\mathtt{2.8}$& $\mathtt{1.5}$& $\mathtt{0.0}$\\
& (7) & IOTA+PTI & $0.68^{+0.06}_{-0.06}$& $\mathtt{51}$& $\mathtt{32}$& $\mathtt{29.2}$& $\mathtt{2.8}$& $\mathtt{1.5}$& $\mathtt{0.0}$\\

\hline 
& & & \multicolumn{7}{c}{Fitted parameters} \\ \cline{4-10} 
K Band & & & $a$ & $\theta$ & $i$ & $f_{c}$  & $f_0$ & $f_1$ & $f_2$\\  
 & &  Obs & [mas] & [$^\circ$] & [$^\circ$] & [\%] & [\%] & $[\%]/\mu\mathrm{m}$ & $[\%]/\mu\mathrm{m}^2$\\ \cline{2-10}   
& (1) & GRAVI21 & $0.61^{+0.01}_{-0.01}$& $45^{+1}_{-1}$& $35^{+1}_{-1}$& $\mathtt{30.0}$& $0.2^{+0.1}_{-0.1}$& $3.6^{+0.1}_{-0.1}$& $0.0^{+0.0}_{-0.0}$\\
& \textbf{(2)} & \textbf{GRAVI17} & $\mathbf{0.66^{+0.04}_{-0.04}}$& $\mathbf{60^{+8}_{-17}}$& $\mathbf{33^{+4}_{-2}}$& $\mathbf{30.0}$& $\mathbf{5.0^{+1.0}_{-1.0}}$& $\mathbf{0.0^{+1.0}_{-1.0}}$& $\mathbf{0.0^{+0.0}_{-0.0}}$\\
& (3) & GRAVI16 & $0.62^{+0.01}_{-0.01}$& $61^{+2}_{-2}$& $35^{+1}_{-1}$& $\mathtt{30.0}$& $2.4^{+0.1}_{-0.1}$& $4.2^{+0.1}_{-0.1}$& $0.0^{+0.0}_{-0.0}$\\
& (4) & CLIMB11 & $0.62^{+0.06}_{-0.05}$& $49^{+3}_{-4}$& $55^{+5}_{-3}$& $29.9^{+1.2}_{-0.5}$& $5.4^{+1.6}_{-1.6}$& $4.0^{+0.1}_{-0.1}$& $0.0^{+0.0}_{-0.0}$\\
& (5) & CLIMB10 & $0.64^{+0.06}_{-0.03}$& $13^{+5}_{-3}$& $45^{+11}_{-15}$& $29.8^{+1.5}_{-0.4}$& $2.2^{+2.9}_{-2.9}$& $1.9^{+0.1}_{-0.1}$& $0.0^{+0.0}_{-0.0}$\\
& (6) & PTI03 & $0.92^{+0.16}_{-0.14}$& $\mathtt{60}$& $\mathtt{33}$& $\mathtt{30.0}$& $\mathtt{5.0}$& $\mathtt{0.0}$& $\mathtt{0.0}$\\
& (7) & VINCI02 & $0.65^{+0.08}_{-0.08}$& $\mathtt{45}$& $\mathtt{35}$& $\mathtt{30.0}$& $\mathtt{0.2}$& $\mathtt{3.6}$& $\mathtt{0.0}$\\
& (8) & PTI00 & $0.80^{+0.05}_{-0.09}$& $\mathtt{60}$& $\mathtt{33}$& $\mathtt{30.0}$& $\mathtt{5.0}$& $\mathtt{0.0}$& $\mathtt{0.0}$\\
& (9) & PTI99 & $0.84^{+0.05}_{-0.08}$& $\mathtt{60}$& $\mathtt{33}$& $\mathtt{30.0}$& $\mathtt{5.0}$& $\mathtt{0.0}$& $\mathtt{0.0}$\\
& (10) & IOTA+PTI & $1.10^{+0.10}_{-0.12}$& $\mathtt{60}$& $\mathtt{33}$& $\mathtt{30.0}$& $\mathtt{5.0}$& $\mathtt{0.0}$& $\mathtt{0.0}$\\

\hline 
\end{tabular}}
\caption{Fitted parameters of the geometrical modeling of FU Orionis. The reference datasets in each band, on which are fixed the sparse dataset (see text), are indicated in bold (values without errorbars).}       
\label{table:results}      
\end{table*}

\newpage
\section{Extended Flux}
\begin{figure*}[!htb]
   \centering
   \includegraphics[width=9cm]{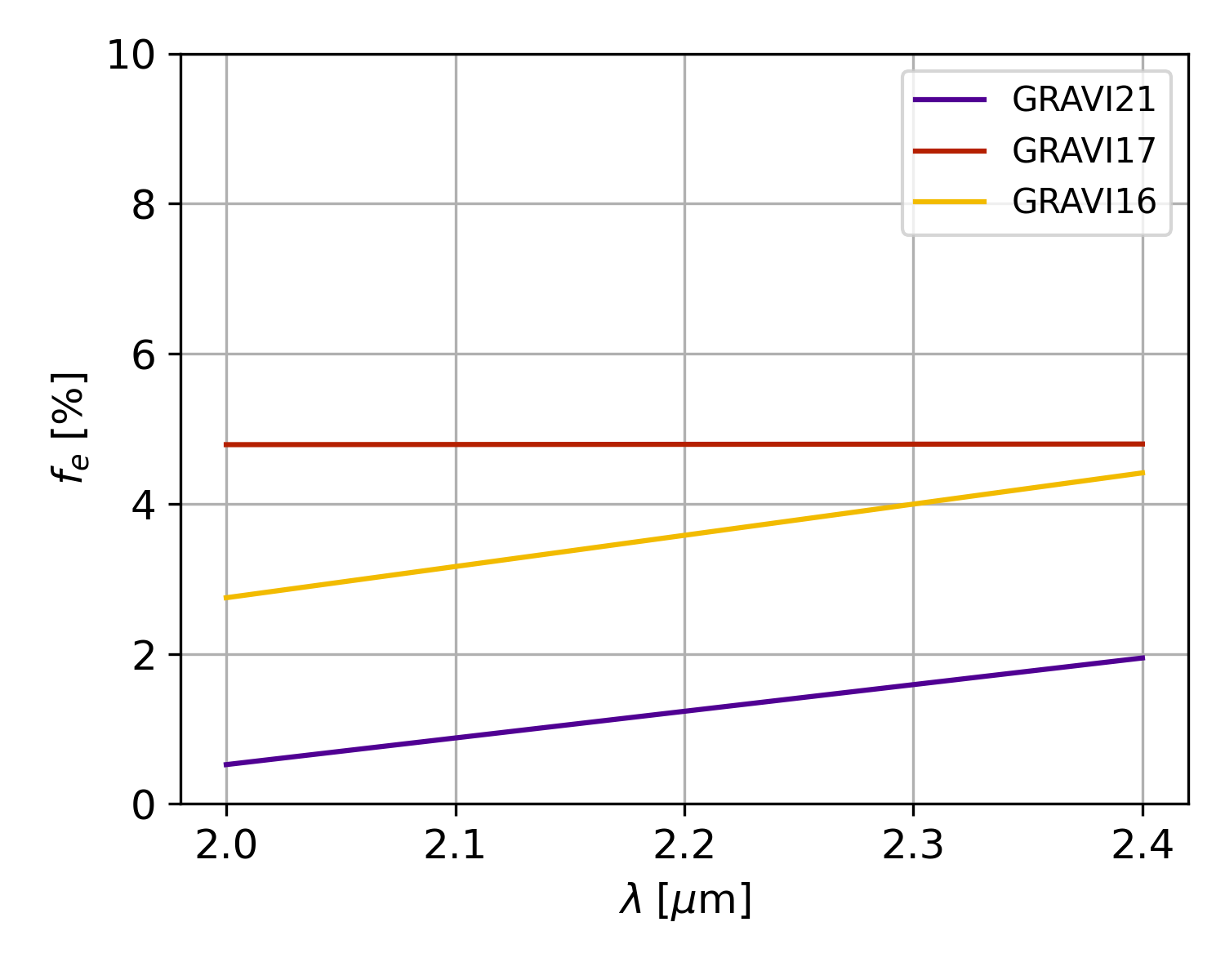}
      \caption{Proportion of the flux of the extended envelope over the total flux measured in the interferometric field of view, in K band. A chromatic dependency is visible in GRAVI21 and GRAVI16 data. The offset between these two curves is due to the smaller interferometric field of view in GRAVI21.}
         \label{fig:envelope}
\end{figure*}



\end{appendix}

\end{document}